\documentclass[conference]{IEEEtran}

\usepackage[skip=1pt,font=small]{caption}
\usepackage{url,graphicx}
\usepackage{balance}
\usepackage{xcolor}
\usepackage{amsmath}
\usepackage[noadjust]{cite}
\usepackage{algpseudocode}
\usepackage{algorithm}
\newcommand{\paragraphbe}[1]{\vspace{0.75ex}\noindent{\bf \em #1} }

\IEEEoverridecommandlockouts
\begin{document}

\title{Membership Inference Attacks Against\\
       Machine Learning Models}
	\author{
		\IEEEauthorblockN{Reza Shokri}
		\IEEEauthorblockA{Cornell Tech\\{\tt shokri@cornell.edu}}
		\and
		\IEEEauthorblockN{Marco Stronati$^{*}$\thanks{$^{*}$This research was performed while the author was at Cornell Tech.}}
		\IEEEauthorblockA{INRIA \\{\tt marco@stronati.org}}
		\and	
		\IEEEauthorblockN{Congzheng Song}
		\IEEEauthorblockA{Cornell\\{\tt cs2296@cornell.edu}}
		\and 
		\IEEEauthorblockN{Vitaly Shmatikov}
		\IEEEauthorblockA{Cornell Tech\\{\tt shmat@cs.cornell.edu}}
	}

\maketitle

\begin{abstract}

We quantitatively investigate how machine learning models leak information
about the individual data records on which they were trained.  We focus on
the basic membership inference attack: given a data record and black-box
access to a model, determine if the record was in the model's training
dataset.  To perform membership inference against a target model, we
make adversarial use of machine learning and train our own inference
model to recognize differences in the target model's predictions on the
inputs that it trained on versus the inputs that it did not train on.

We empirically evaluate our inference techniques on classification models
trained by commercial ``machine learning as a service'' providers such
as Google and Amazon.  Using realistic datasets and classification tasks,
including a hospital discharge dataset whose membership is sensitive from
the privacy perspective, we show that these models can be vulnerable
to membership inference attacks.  We then investigate the factors that
influence this leakage and evaluate mitigation strategies.

\end{abstract}

\section{Introduction}

Machine learning is the foundation of popular Internet services such
as image and speech recognition and natural language translation.
Many companies also use machine learning internally, to improve marketing
and advertising, recommend products and services to users, or better
understand the data generated by their operations.  In all of these
scenarios, activities of individual users\textemdash their purchases
and preferences, health data, online and offline transactions, photos
they take, commands they speak into their mobile phones, locations they
travel to\textemdash are used as the training data.

Internet giants such as Google and Amazon are already offering ``machine
learning as a service.''  Any customer in possession of a dataset and a
data classification task can upload this dataset to the service and pay
it to construct a model.  The service then makes the model available to
the customer, typically as a black-box API.  For example, a mobile-app
maker can use such a service to analyze users' activities and query the
resulting model inside the app to promote in-app purchases to users when
they are most likely to respond.  Some machine-learning services also
let data owners expose their models to external users for querying or
even sell them.

\paragraphbe{Our contributions.}
We focus on the fundamental question known as \textbf{membership
inference}: given a machine learning model and a record, determine whether
this record was used as part of the model's training dataset or not.
We investigate this question in the most difficult setting, where
the adversary's access to the model is limited to \textbf{black-box}
queries that return the model's output on a given input.  In summary,
we quantify membership information leakage through the prediction outputs
of machine learning models.

To answer the membership inference question, we turn machine learning
against itself and train an \emph{attack model} whose purpose is to
distinguish the target model's behavior on the training inputs from
its behavior on the inputs that it did not encounter during training.
In other words, we turn the membership inference problem into a
classification problem.

Attacking black-box models such as those built by commercial ``machine
learning as a service'' providers requires more sophistication than
attacking white-box models whose structure and parameters are known
to the adversary.  To construct our attack models, we invented a
\textbf{shadow training} technique.  First, we create multiple ``shadow
models'' that imitate the behavior of the target model, but for which
we know the training datasets and thus the ground truth about membership
in these datasets.  We then train the attack model on the labeled inputs
and outputs of the shadow models.

We developed several effective methods to generate training data for
the shadow models.  The first method uses black-box access to the target
model to synthesize this data.  The second method uses statistics about
the population from which the target's training dataset was drawn.
The third method assumes that the adversary has access to a potentially
noisy version of the target's training dataset.  The first method does
not assume any prior knowledge about the distribution of the target
model's training data, while the second and third methods allow the
attacker to query the target model only \emph{once} before inferring
whether a given record was in its training dataset.

Our inference techniques are generic and not based on any particular
dataset or model type.  We evaluate them against neural networks, as well
as black-box models trained using Amazon ML and Google Prediction API.
All of our experiments on Amazon's and Google's platforms were done
without knowing the learning algorithms used by these services, nor
the architecture of the resulting models, since Amazon and Google don't
reveal this information to the customers.  For our evaluation, we use
realistic classification tasks and standard model-training procedures
on concrete datasets of images, retail purchases, location traces, and
hospital inpatient stays.  In addition to demonstrating that membership
inference attacks are successful, we quantify how their success relates
to the classification tasks and the standard metrics of overfitting.

Inferring information about the model's training dataset should
not be confused with techniques such as model inversion that use
a model's output on a hidden input to infer something about this
input~\cite{fredrikson2014privacy} or to extract features that
characterize one of the model's classes~\cite{fredrikson2015model}.
As explained in~\cite{frankblog} and Section~\ref{sec:relatedwork},
model inversion does not produce an actual member of the model's training
dataset, nor, given a record, does it infer whether this record was in
the training dataset.  By contrast, the membership inference problem
we study in this paper is essentially the same as the well-known
problem of identifying the presence of an individual's data in a mixed
pool given some statistics about the pool~\cite{homer2008resolving,
sankararaman2009genomic, dwork2015robust, backes2016membership}.  In our
case, however, the goal is to infer membership given a black-box API to
a model of unknown structure, as opposed to explicit statistics.

Our experimental results show that models created using
machine-learning-as-a-service platforms can leak a lot of information
about their training datasets.  For multi-class classification models
trained on 10,000-record retail transaction datasets using Google's and
Amazon's services in default configurations, our membership inference
achieves median accuracy of $94\%$ and $74\%$, respectively.  Even if we
make no prior assumptions about the distribution of the target model's
training data and use fully synthetic data for our shadow models, the
accuracy of membership inference against Google-trained models is $90\%$.
Our results for the Texas hospital discharge dataset (over 70\% accuracy)
indicate that membership inference can present a risk to health-care
datasets if these datasets are used to train machine learning models
and access to the resulting models is open to the public.  Membership in
such datasets is highly sensitive.

We discuss the root causes that make these attacks possible and
quantitatively compare mitigation strategies such as limiting the model's
predictions to top $k$ classes, decreasing the precision of the prediction
vector, increasing its entropy, or using regularization while training
the model.

In summary, this paper demonstrates and quantifies the problem of machine
learning models leaking information about their training datasets.
To create our attack models, we developed a new shadow learning technique
that works with minimal knowledge about the target model and its training
dataset.  Finally, we quantify how the leakage of membership information
is related to model overfitting.

\section{Machine Learning Background}

Machine learning algorithms help us better understand and analyze complex
data.  When the model is created using \emph{unsupervised} training, the
objective is to extract useful features from the unlabeled data and build
a model that explains its hidden structure.  When the model is created
using \emph{supervised} training, which is the focus of this paper,
the training records (as inputs of the model) are assigned labels or
scores (as outputs of the model).  The goal is to learn the relationship
between the data and the labels and construct a model that can generalize
to data records beyond the training set~\cite{hastie2005elements}.
Model-training algorithms aim to minimize the model's prediction
error on the training dataset and thus may overfit to this dataset,
producing models that perform better on the training inputs than on the
inputs drawn from the same population but not used during the training.
Many \emph{regularization} techniques have been proposed to prevent models
from becoming overfitted to their training datasets while minimizing
their prediction error~\cite{hastie2005elements}.

Supervised training is often used for classification and other prediction
tasks.  For example, a retailer may train a model that predicts a
customer's shopping style in order to offer her suitable incentives, while
a medical researcher may train a model to predict which treatment is most
likely to succeed given a patient's clinical symptoms or genetic makeup.

\paragraphbe{Machine learning as a service.}
Major Internet companies now offer machine learning as a service
on their cloud platforms.  Examples include Google Prediction
API,\footnote{\url{https://cloud.google.com/prediction}}
Amazon Machine Learning (Amazon
ML),\footnote{\url{https://aws.amazon.com/machine-learning}}
Microsoft Azure Machine Learning (Azure
ML),\footnote{\url{https://studio.azureml.net}} and
BigML.\footnote{\url{https://bigml.com}}

These platforms provide simple APIs for uploading the data and
for training and querying models, thus making machine learning
technologies available to any customer.  For example, a developer
may create an app that gathers data from users, uploads it into
the cloud platform to train a model (or update an existing model
with new data), and then uses the model's predictions inside the
app to improve its features or better interact with the users.
Some platforms even envision data holders training a model
and then sharing it with others through the platform's API for
profit.\footnote{\url{https://cloud.google.com/prediction/docs/gallery}}

The details of the models and the training algorithms are hidden from
the data owners.  The type of the model may be chosen by the service
adaptively, depending on the data and perhaps accuracy on validation
subsets.  Service providers do not warn customers about the consequences
of overfitting and provide little or no control over regularization.
For example, Google Prediction API hides all details, while Amazon ML
provides only a very limited set of pre-defined options (L1- or L2-norm
regularization).  The models cannot be downloaded and are accessed only
through the service's API.  Service providers derive revenue mainly by
charging customers for queries through this API.  Therefore, we treat
``machine learning as a service'' as a black box.  All inference attacks
we demonstrate in this paper are performed entirely through the services'
standard APIs.

\section{Privacy in Machine Learning}
\label{meaning}

Before dealing with inference attacks, we need to define what privacy
means in the context of machine learning or, alternatively, what it
means for a machine learning model to breach privacy.

\subsection{Inference about members of the population}

A plausible notion of privacy, known in statistical disclosure control
as the ``Dalenius desideratum,'' states that the model should reveal no
more about the input to which it is applied than would have been known
about this input without applying the model.  This cannot be achieved
by any useful model~\cite{dworknaor}.

A related notion of privacy appears in prior work on model
inversion~\cite{fredrikson2014privacy}: a privacy breach occurs if an
adversary can use the model's output to infer the values of unintended
(sensitive) attributes used as input to the model.  As observed
in~\cite{frankblog}, it may not be possible to prevent this ``breach''
if the model is based on statistical facts about the population.
For example, suppose that training the model has uncovered a high
correlation between a person's externally observable phenotype features
and their genetic predisposition to a certain disease.  This correlation
is now a publicly known scientific fact that allows anyone to infer
information about the person's genome after observing that person.

Critically, this correlation applies to \emph{all} members of a given
population.  Therefore, the model breaches ``privacy'' not just of the
people whose data was used to create the model, but also of other people
from the same population, even those whose data was not used and whose
identities may not even be known to the model's creator (i.e., this
is ``spooky action at a distance'').  Valid models generalize, i.e.,
they make accurate predictions on inputs that were not part of their
training datasets.  This means that the creator of a generalizable model
cannot do anything to protect ``privacy'' as defined above because the
correlations on which the model is based\textemdash and the inferences
that these correlations enable\textemdash hold for the entire population,
regardless of how the training sample was chosen or how the model was
created from this sample.

\subsection{Inference about members of the training dataset}

To bypass the difficulties inherent in defining and protecting
privacy of the entire population, we focus on protecting privacy
of the individuals whose data was used to train the model.  This
motivation is closely related to the original goals of differential
privacy~\cite{dwork2006calibrating}.

Of course, members of the training dataset are members of the population,
too.  We investigate what the model reveals about them \emph{beyond} what
it reveals about an arbitrary member of the population.  Our ultimate
goal is to measure the \emph{membership risk} that a person incurs if
they allow their data to be used to train a model.

The basic attack in this setting is \textbf{membership inference},
i.e., determining whether a given data record was part of the model's
training dataset or not.  When a record is fully known to the adversary,
learning that it was used to train a particular model is an indication
of information leakage through the model.  In some cases, it can directly
lead to a privacy breach.  For example, knowing that a certain patient's
clinical record was used to train a model associated with a disease (e.g,
to determine the appropriate medicine dosage or to discover the genetic
basis of the disease) can reveal that the patient has this disease.

We investigate the membership inference problem in the black-box scenario
where the adversary can only supply inputs to the model and receive
the model's output(s).  In some situations, the model is available
to the adversary indirectly.  For example, an app developer may use a
machine-learning service to construct a model from the data collected
by the app and have the app make API calls to the resulting model.
In this case, the adversary would supply inputs to the app (rather
than directly to the model) and receive the app's outputs (which are
based on the model's outputs).  The details of internal model usage
vary significantly from app to app.  For simplicity and generality, we
will assume that the adversary directly supplies inputs to and receives
outputs from the black-box model.

\section{Problem Statement}
\label{sec:problem}

Consider a set of labeled data records sampled from some population and
partitioned into classes.  We assume that a machine learning algorithm
is used to train a classification model that captures the relationship
between the content of the data records and their labels.  

For any input data record, the model outputs the \emph{prediction vector}
of probabilities, one per class, that the record belongs to a certain
class.  We will also refer to these probabilities as \emph{confidence
values}.  The class with the highest confidence value is selected as
the predicted label for the data record.  The accuracy of the model is
evaluated by measuring how it generalizes beyond its training set and
predicts the labels of other data records from the same population.

We assume that the attacker has query access to the model and can obtain
the model's prediction vector on any data record.  The attacker knows
the format of the inputs and outputs of the model, including their
number and the range of values they can take.  We also assume that
the attacker either (1) knows the type and architecture of the machine
learning model, as well as the training algorithm, or (2) has black-box
access to a machine learning oracle (e.g., a ``machine learning as a
service'' platform) that was used to train the model.  In the latter
case, the attacker does \emph{not} know a priori the model's structure
or meta-parameters.

The attacker may have some background knowledge about the population
from which the target model's training dataset was drawn.  For example,
he may have independently drawn samples from the population, disjoint
from the target model's training dataset.  Alternatively, the attacker
may know some general statistics about the population, for example,
the marginal distribution of feature values.

The setting for our inference attack is as follows.  The attacker is given
a data record and black-box query access to the target model.  The attack
succeeds if the attacker can correctly determine whether this data record
was part of the model's training dataset or not.  The standard metrics for
attack accuracy are \emph{precision} (what fraction of records inferred
as members are indeed members of the training dataset) and \emph{recall}
(what fraction of the training dataset's members are correctly inferred
as members by the attacker).

\section{Membership Inference}
\label{sec:attacks}

\begin{figure}[t!]
\centering
\includegraphics[width=0.95\columnwidth]{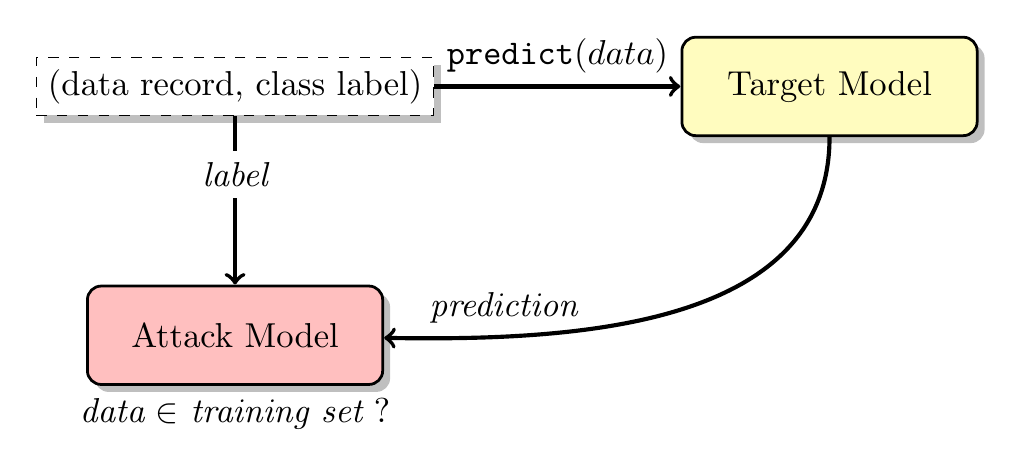}
\caption{\small Membership inference attack in the black-box setting.
The attacker queries the target model with a data record and obtains
the model's prediction on that record.  The prediction is a vector of
probabilities, one per class, that the record belongs to a certain class.
This prediction vector, along with the label of the target record, is
passed to the attack model, which infers whether the record was {\em in}
or {\em out} of the target model's training dataset.}\label{fig:attack}
\end{figure}

\begin{figure}[t!]
\centering
\includegraphics[width=0.95\columnwidth]{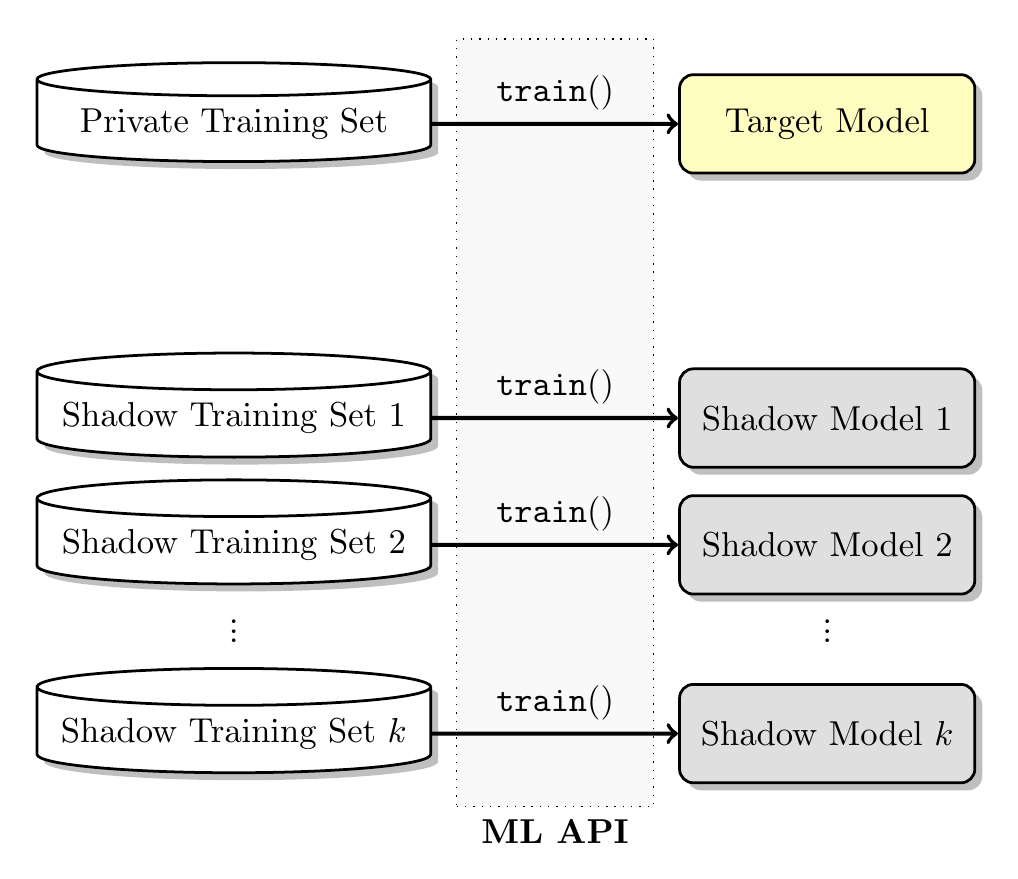}
\caption{\small Training shadow models using the same machine learning
platform as was used to train the target model.  The training datasets
of the target and shadow models have the same format but are disjoint.
The training datasets of the shadow models may overlap.  All models'
internal parameters are trained independently.}\label{fig:target_shadow}
\end{figure}

\subsection{Overview of the attack}

Our membership inference attack exploits the observation that
machine learning models often behave differently on the data that
they were trained on versus the data that they ``see'' for the
first time.  Overfitting is a common reason but not the only one (see
Section~\ref{sec:whyattackworks}).  The objective of the attacker is
to construct an \emph{attack model} that can recognize such differences
in the target model's behavior and use them to distinguish members from
non-members of the target model's training dataset based solely on the
target model's output.

Our attack model is a collection of models, one for each output class
of the target model.  This increases accuracy of the attack because the
target model produces different distributions over its output classes
depending on the input's true class.

To train our attack model, we build multiple ``shadow'' models intended
to behave similarly to the target model.  In contrast to the target
model, we know the ground truth for each shadow model, i.e., whether a
given record was in its training dataset or not.  Therefore, we can use
supervised training on the inputs and the corresponding outputs (each
labeled ``in'' or ``out'') of the shadow models to teach the attack
model how to distinguish the shadow models' outputs on members of their
training datasets from their outputs on non-members.

Formally, let $f_{\mathsf{target}}()$ be the target model,
and let $D_{\mathsf{target}}^{\mathsf{train}}$ be its
private training dataset which contains labeled data records
$(\mathbf{x}^{\{i\}}, y^{\{i\}})_{\mathsf{target}}$.  A data record
$\mathbf{x}^{\{i\}}_{\mathsf{target}}$ is the input to the model, and
$y^{\{i\}}_{\mathsf{target}}$ is the true label that can take values
from a set of classes of size $c_{\mathsf{target}}$.  The output of
the target model is a probability vector of size $c_{\mathsf{target}}$.
The elements of this vector are in $[0,1]$ and sum up to $1$.

Let $f_{\mathsf{attack}}()$ be the attack model.  Its input
$\mathbf{x}_{\mathsf{attack}}$ is composed of a correctly labeled record
and a prediction vector of size $c_{\mathsf{target}}$.  Since the goal
of the attack is decisional membership inference, the attack model is
a binary classifier with two output classes, ``in'' and ``out.''

Figure~\ref{fig:attack} illustrates our end-to-end attack process.  For a
labeled record $(\mathbf{x}, y)$, we use the target model to compute
the prediction vector $\mathbf{y} = f_{\mathsf{target}}(\mathbf{x})$.
The distribution of $\mathbf{y}$ (classification confidence values)
depends heavily on the true class of $\mathbf{x}$.  This is why we
pass the true label $y$ of $\mathbf{x}$ in addition to the model's
prediction vector $\mathbf{y}$ to the attack model.  Given how the
probabilities in $\mathbf{y}$ are distributed around $y$, the attack
model computes the membership probability $\Pr\{(\mathbf{x}, y)
\in D_{\mathsf{target}}^{\mathsf{train}}\}$, i.e., the probability
that $((\mathbf{x}, y), \mathbf{y})$ belongs to the ``in'' class
or, equivalently, that $\mathbf{x}$ is in the training dataset of
$f_{\mathsf{target}}()$.

The main challenge is how to train the attack model to distinguish
members from non-members of the target model's training dataset when
the attacker has no information about the internal parameters of the
target model and only limited query access to it through the public API.
To solve this conundrum, we developed a \emph{shadow training} technique
that lets us train the attack model on proxy targets for which we do
know the training dataset and can thus perform supervised training.

\subsection{Shadow models}

The attacker creates $k$ shadow models $f_{\mathsf{shadow}}^{\,i}()$.
Each shadow model $i$ is trained on a dataset
$D_{\mathsf{shadow}^{\,i}}^{\mathsf{train}}$ of the same format
as and distributed similarly to the target model's training
dataset.  These shadow training datasets can be generated
using one of methods described in Section~\ref{shadowtrain}.
We assume that the datasets used for training the shadow models
are disjoint from the private dataset used to train the target
model ($\forall i, D_{\mathsf{shadow}^{\,i}}^{\mathsf{train}} \cap
D_{\mathsf{target}}^{\mathsf{train}} = \emptyset$).  This is the worst
case for the attacker; the attack will perform even better if the training
datasets happen to overlap.

The shadow models must be trained in a similar way to the target model.
This is easy if the target's training algorithm (e.g., neural networks,
SVM, logistic regression) and model structure (e.g., the wiring of
a neural network) are known.  Machine learning as a service is more
challenging.  Here the type and structure of the target model are not
known, but the attacker can use exactly the same service (e.g., Google
Prediction API) to train the shadow model as was used to train the target
model\textemdash see Figure~\ref{fig:target_shadow}.

The more shadow models, the more accurate the attack model will be.
As described in Section~\ref{attacktrain}, the attack model is trained
to recognize differences in shadow models' behavior when these models
operate on inputs from their own training datasets versus inputs they
did not encounter during training.  Therefore, more shadow models provide
more training fodder for the attack model.

\algnewcommand{\Cmnt}[1]{\Comment{\textcolor{gray}{\small\em #1}}}
\begin{algorithm}[t]
  \caption{Data synthesis using the target model}\label{synthesis}
  \begin{algorithmic}[1]
  \Procedure{Synthesize}{$\mathrm{class}: c$}
    \State $\mathbf{x} \gets$ \Call{RandRecord}{\textcolor{white}{.}} \Cmnt{initialize a record randomly}
    \State $y_c^* \gets 0$
    \State $j \gets 0$
    \State $k \gets k_{max}$
    
    \For{$iteration = 1 \cdots iter_{max}$}
      \State $\mathbf{y} \gets f_{\mathsf{target}}(\mathbf{x})$  \Cmnt{query the target model}
      \If{$y_c \ge y_c^*$} \Cmnt{accept the record}
        \If{$y_c > \mathrm{conf}_{min}$ and $c = \arg\max(\mathbf{y})$} 
          \If{$\mathrm{rand}() < y_c$} \Cmnt{sample}
            \State \textbf{return} $\mathbf{x}$ \Cmnt{synthetic data}
          \EndIf
        \EndIf
        \State $\mathbf{x}^* \gets \mathbf{x}$
        \State $y_c^* \gets y_c$
        \State $j \gets 0$
      \Else 
        \State $j \gets j+1$
        \If{$j > rej_{max}$} \Cmnt{many consecutive rejects}
          \State $k \gets \max(k_{min}, \lceil k/2 \rceil)$
          \State $j \gets 0$
        \EndIf
      \EndIf
      \State $\mathbf{x} \gets$ \Call{RandRecord}{$\mathbf{x}^*$, $k$} \Cmnt{randomize $k$ features}
    \EndFor

    \State \textbf{return} $\bot$ \Cmnt{failed to synthesize}
  \EndProcedure
\end{algorithmic}
\end{algorithm}

\begin{figure*}[t!]
\centering
\includegraphics[width=1.8\columnwidth]{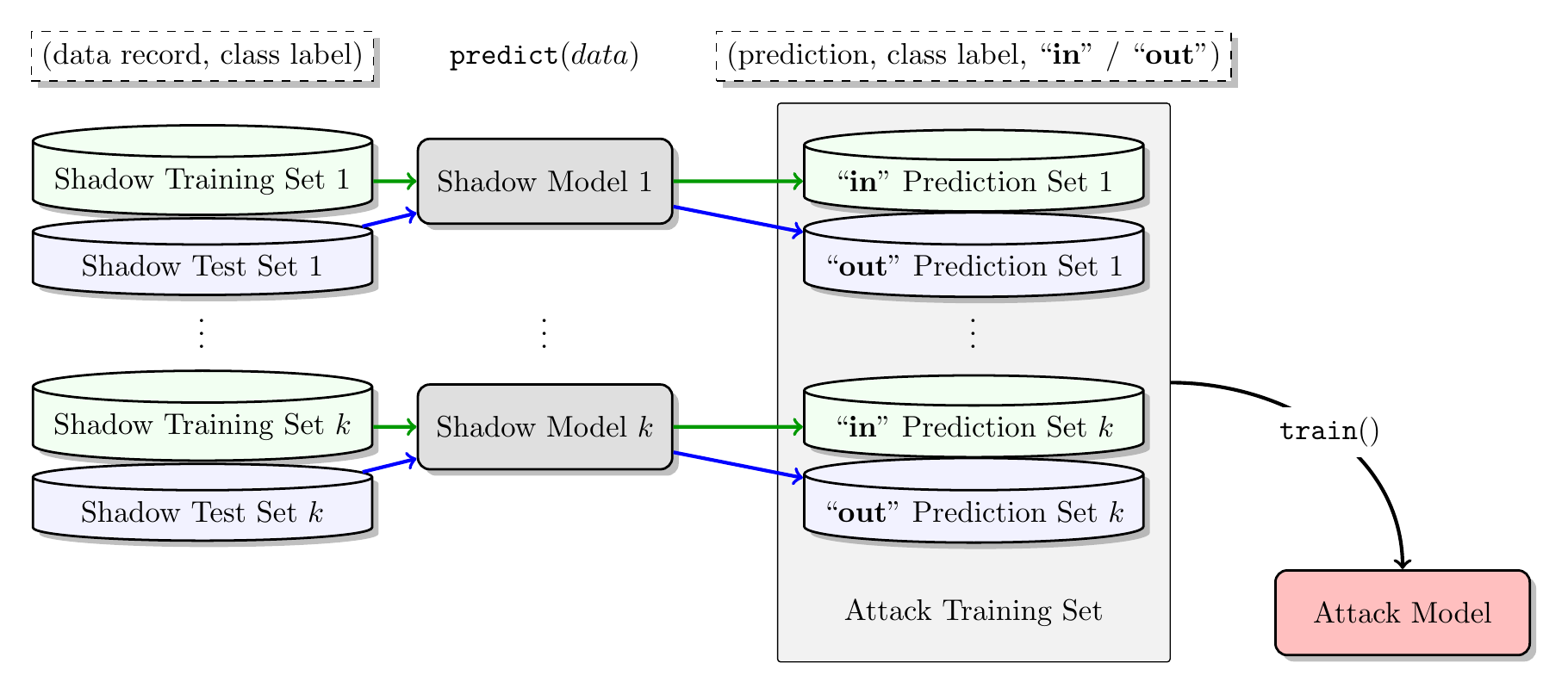}
\caption{\small Training the attack model on the inputs and outputs of
the shadow models.  For all records in the training dataset of a shadow
model, we query the model and obtain the output.  These output vectors
are labeled ``in'' and added to the attack model's training dataset.
We also query the shadow model with a test dataset disjoint from its
training dataset.  The outputs on this set are labeled ``out'' and also
added to the attack model's training dataset.  Having constructed a
dataset that reflects the black-box behavior of the shadow models
on their training and test datasets, we train a collection of
$c_{\mathsf{target}}$ attack models, one per each output class of the
target model.}\label{fig:shadows_attack}
\end{figure*}

\subsection{Generating training data for shadow models}
\label{shadowtrain}

To train shadow models, the attacker needs training data that is
distributed similarly to the target model's training data.  We developed
several methods for generating such data.

\paragraphbe{Model-based synthesis.}
If the attacker does not have real training data nor any statistics about
its distribution, he can generate synthetic training data for the shadow
models using the target model itself.  The intuition is that records
that are classified by the target model with high confidence should be
statistically similar to the target's training dataset and thus provide
good fodder for shadow models.

The synthesis process runs in two phases: (1) \emph{search}, using a
hill-climbing algorithm, the space of possible data records to find
inputs that are classified by the target model with high confidence;
(2) \emph{sample} synthetic data from these records.  After this process
synthesizes a record, the attacker can repeat it until the training
dataset for shadow models is full.

See Algorithm~1 for the pseudocode of our synthesis procedure.  First,
fix class $c$ for which the attacker wants to generate synthetic data.
The first phase is an iterative process.  Start by randomly initializing
a data record $\mathbf{x}$.  Assuming that the attacker knows only the
syntactic format of data records, sample the value for each feature
uniformly at random from among all possible values of that feature.
In each iteration, propose a new record.  A proposed record is
\emph{accepted} only if it increases the hill-climbing objective: the
probability of being classified by the target model as class $c$.

Each iteration involves proposing a new candidate record by changing $k$
randomly selected features of the latest accepted record $\mathbf{x}^*$.
This is done by flipping binary features or resampling new values
for features of other types.  We initialize $k$ to $k_{max}$ and
divide it by $2$ when $rej_{max}$ subsequent proposals are rejected.
This controls the diameter of search around the accepted record in order
to propose a new record.  We set the minimum value of $k$ to $k_{min}$.
This controls the speed of the search for new records with a potentially
higher classification probability $y_c$.

The second, sampling phase starts when the target model's probability
$y_c$ that the proposed data record is classified as belonging to
class $c$ is larger than the probabilities for all other classes and
also larger than a threshold $\mathrm{conf}_{min}$.  This ensures that
the predicted label for the record is $c$, and that the target model is
sufficiently confident in its label prediction.  We select such record for
the synthetic dataset with probability $y_c^*$ and, if selection fails,
repeat until a record is selected.

This synthesis procedure works only if the adversary can efficiently
explore the space of possible inputs and discover inputs that are
classified by the target model with high confidence.  For example, it
may not work if the inputs are high-resolution images and the target
model performs a complex image classification task.

\paragraphbe{Statistics-based synthesis.}
The attacker may have some statistical information about the population
from which the target model's training data was drawn.  For example,
the attacker may have prior knowledge of the marginal distributions of
different features.  In our experiments, we generate synthetic training
records for the shadow models by independently sampling the value of
each feature from its own marginal distribution.  The resulting attack
models are very effective.

\paragraphbe{Noisy real data.}
The attacker may have access to some data that is similar to the
target model's training data and can be considered as a ``noisy''
version thereof.  In our experiments with location datasets, we simulate
this by flipping the (binary) values of 10\% or 20\% randomly selected
features, then training our shadow models on the resulting noisy dataset.
This scenario models the case where the training data for the target
and shadow models are not sampled from exactly the same population,
or else sampled in a non-uniform way.

\subsection{Training the attack model}
\label{attacktrain}

The main idea behind our shadow training technique is that similar models
trained on relatively similar data records using the same service behave
in a similar way.  This observation is empirically borne out by our
experiments in the rest of this paper.  Our results show that learning
how to infer membership in shadow models' training datasets (for which
we know the ground truth and can easily compute the cost function during
supervised training) produces an attack model that successfully infers
membership in the target model's training dataset, too.

We query each shadow model with its own training dataset and with a
disjoint test set of the same size.  The outputs on the training dataset
are labeled ``in,'' the rest are labeled ``out.''  Now, the attacker has
a dataset of records, the corresponding outputs of the shadow models,
and the in/out labels.  The objective of the attack model is to infer
the labels from the records and corresponding outputs.

Figure~\ref{fig:shadows_attack} shows how to
train the attack model.  For all $(\mathbf{x}, y) \in
D_{\mathsf{shadow}^{\,i}}^{\mathsf{train}}$, compute the prediction
vector $\mathbf{y} = f_{\mathsf{shadow}}^{\,i}(\mathbf{x})$
and add the record $(y, \mathbf{y}, \mathsf{in})$ to the
attack training set $D_{\mathsf{attack}}^{\mathsf{train}}$.
Let $D_{\mathsf{shadow}^{\,i}}^{\mathsf{test}}$ be a set of records
disjoint from the training set of the $i$th shadow model. Then, $\forall
(\mathbf{x}, y) \in D_{\mathsf{shadow}^{\,i}}^{\mathsf{test}}$ compute the
prediction vector $\mathbf{y} = f_{\mathsf{shadow}}^{\,i}(\mathbf{x})$
and add the record $(y, \mathbf{y}, \mathsf{out})$ to the attack
training set $D_{\mathsf{attack}}^{\mathsf{train}}$.  Finally, split
$D_{\mathsf{attack}}^{\mathsf{train}}$ into $c_{target}$ partitions,
each associated with a different class label.  For each label $y$, train
a separate model that, given $\mathbf{y}$, predicts the $\mathsf{in}$
or $\mathsf{out}$ membership status for $\mathbf{x}$.

If we use model-based synthesis from Section~\ref{shadowtrain}, all of
the raw training data for the attack model is drawn from the records
that are classified by the target model with high confidence.  This is
true, however, both for the records used \emph{in} the shadow models'
training datasets and for the test records left \emph{out} of these
datasets.  Therefore, it is not the case that the attack model simply
learns to recognize inputs that are classified with high confidence.
Instead, it learns to perform a much subtler task: how to distinguish
between the training inputs classified with high confidence and other,
non-training inputs that are also classified with high confidence.

In effect, we convert the problem of recognizing the complex relationship
between members of the training dataset and the model's output into a
binary classification problem.  Binary classification is a standard
machine learning task, thus we can use any state-of-the-art machine
learning framework or service to build the attack model.  Our approach
is independent of the specific method used for attack model training.
For example, in Section~\ref{sec:evaluation} we construct the attack model
using neural networks and also using the same black-box Google Prediction
API that we are attacking, in which case we have no control over the model
structure, model parameters, or training meta-parameters\textemdash but
still obtain a working attack model.

\section{Evaluation}
\label{sec:evaluation}

We first describe the datasets that we use for evaluation, followed
by the description of the target models and our experimental setup.
We then present the results of our membership inference attacks in
several settings and study in detail how and why the attacks work against
different datasets and machine learning platforms.

\subsection{Data}
\label{sec:data}

\paragraphbe{CIFAR.}
CIFAR-10 and CIFAR-100 are benchmark datasets used to evaluate image
recognition algorithms~\cite{krizhevsky2009learning}.  CIFAR-10 is
composed of $32\times32$ color images in $10$ classes, with $6,000$
images per class.  In total, there are $50,000$ training images and
$10,000$ test images.  CIFAR-100 has the same format as CIFAR-10,
but it has $100$ classes containing $600$ images each. There are $500$
training images and $100$ testing images per class.  We use different
fractions of this dataset in our attack experiments to show the effect
of the training dataset size on the accuracy of the attack.

\paragraphbe{Purchases.}
Our purchase dataset is based on Kaggle's ``acquire valued shoppers''
challenge dataset that contains shopping histories for several thousand
individuals.\footnote{\url{https://kaggle.com/c/acquire-valued-shoppers-challenge/data}}
The purpose of the challenge is to design accurate coupon promotion
strategies.  Each user record contains his or her transactions over
a year.  The transactions include many fields such as product name,
store chain, quantity, and date of purchase.

For our experiments, we derived a simplified purchase dataset (with
$197,324$ records), where each record consists of $600$ binary features.
Each feature corresponds to a product and represents whether the
user has purchased it or not.  To design our classification tasks,
we first cluster the records into multiple classes, each representing
a different purchase style.  In our experiments, we use $5$ different
classification tasks with a different number of classes $\{2, 10, 20,
50, 100\}$.  The classification task is to predict the purchase style
of a user given the $600$-feature vector.  We use $10,000$ randomly
selected records from the purchase dataset to train the target model.
The rest of the dataset contributes to the test set and (if necessary)
the training sets of the shadow models.

\paragraphbe{Locations.}
\label{locations}
We created a location dataset from the publicly available set of
mobile users' location ``check-ins'' in the Foursquare social network,
restricted to the Bangkok area and collected from April 2012 to September
2013~\cite{yang2016participatory}.\footnote{\url{https://sites.google.com/site/yangdingqi/home/foursquare-dataset}}
The check-in dataset contains $11,592$ users and $119,744$ locations, for
a total of $1,136,481$ check-ins.  We filtered out users with fewer than
25 check-ins and venues with fewer than 100 visits, which left us with
$5,010$ user profiles.  For each location venue, we have the geographical
position as well as its location type (e.g., Indian restaurant, fast
food, etc.).  The total number of location types is $128$.  We partition
the Bangkok map into areas of size $0.5km \times 0.5km$, yielding $318$
regions for which we have at least one user check-in.

Each record in the resulting dataset has $446$ binary features,
representing whether the user visited a certain region or location type,
i.e., the user's semantic and geographical profile.  The classification
task is similar to the purchase dataset.  We cluster the location
dataset into $30$ classes, each representing a different geosocial type.
The classification task is to predict the user's geosocial type given
his or her record.  We use $1,600$ randomly selected records to train
the target model.  The rest of the dataset contributes to the test set
and (if necessary) the training sets of the shadow models.

\paragraphbe{Texas hospital stays.}
This dataset is based on the Hospital Discharge Data public
use files with information about inpatients stays in several health
facilities,\footnote{\url{https://www.dshs.texas.gov/THCIC/Hospitals/Download.shtm}}
released by the Texas Department of State Health Services from 2006 to
2009.  Each record contains four main groups of attributes: the external
causes of injury (e.g., suicide, drug misuse), the diagnosis (e.g.,
schizophrenia, illegal abortion), the procedures the patient underwent
(e.g., surgery) and some generic information such as the gender, age,
race, hospital id, and length of stay.

Our classification task is to predict the patient's main procedure based
on the attributes other than secondary procedures.  We focus on the $100$
most frequent procedures.  The resulting dataset has $67,330$ records
and $6,170$ binary features. We use $10,000$ randomly selected records
to train the target model.

Note that our experiments do not involve re-identification of known
individuals and fully comply with the data use agreement for the original
Public Use Data File.

\paragraphbe{MNIST.}
This is a dataset of $70,000$ handwritten
digits formatted as $32\times32$
images and normalized so that the digits are located at the center
of the image.\footnote{\url{http://yann.lecun.com/exdb/mnist}}  We use $10,000$ randomly selected images to train the
target model.

\paragraphbe{UCI Adult (Census Income).}
This dataset includes $48,842$ records with $14$ attributes such
as age, gender, education, marital status, occupation, working
hours, and native country.  The (binary) classification task is
to predict if a person makes over \$50K a year based on the census
attributes.\footnote{\url{http://archive.ics.uci.edu/ml/datasets/Adult}} We
use $10,000$ randomly selected records to train the target model.

\begin{figure*}[t]
\centering
\includegraphics[width=0.65\columnwidth]{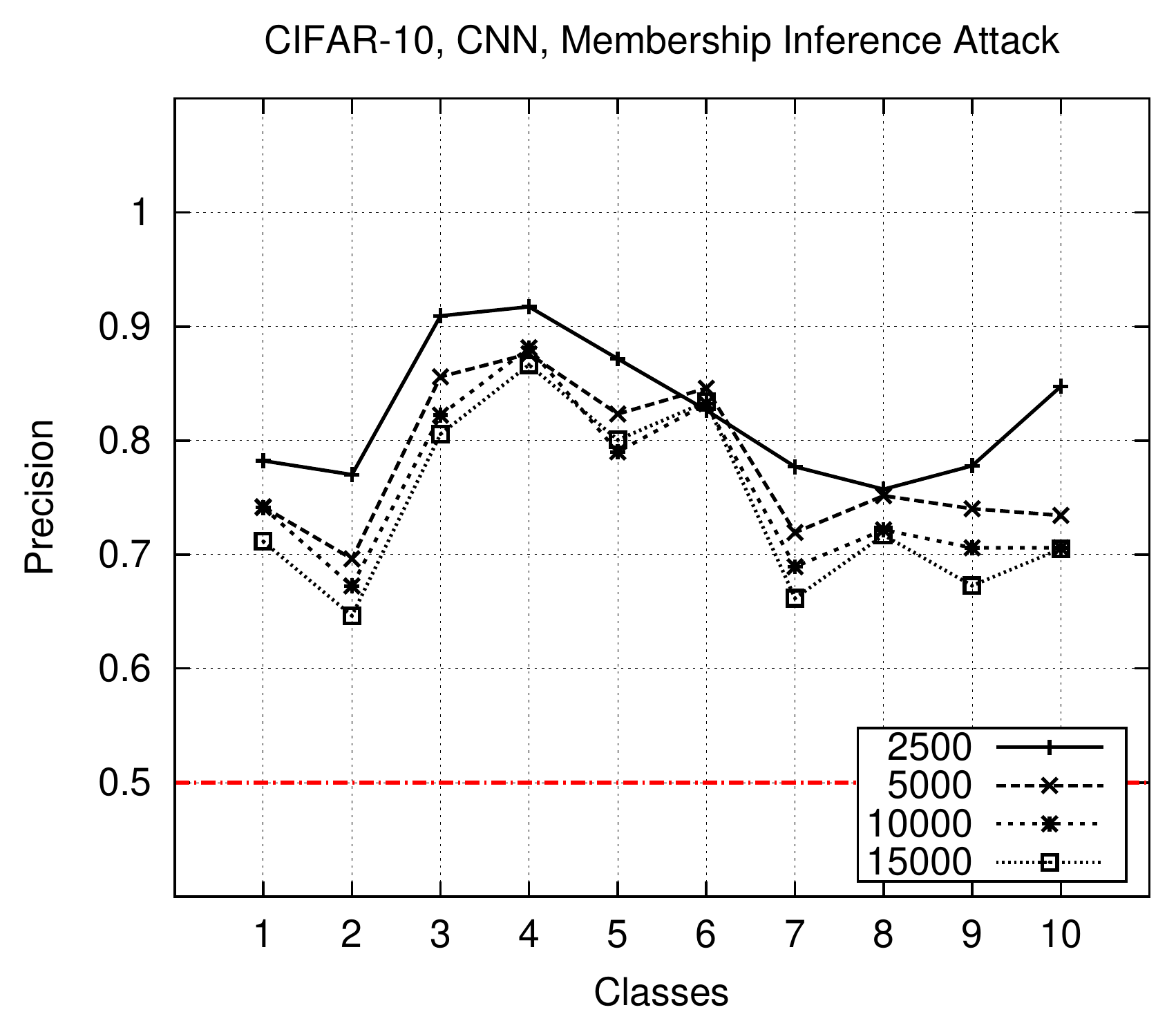}
\includegraphics[width=0.65\columnwidth]{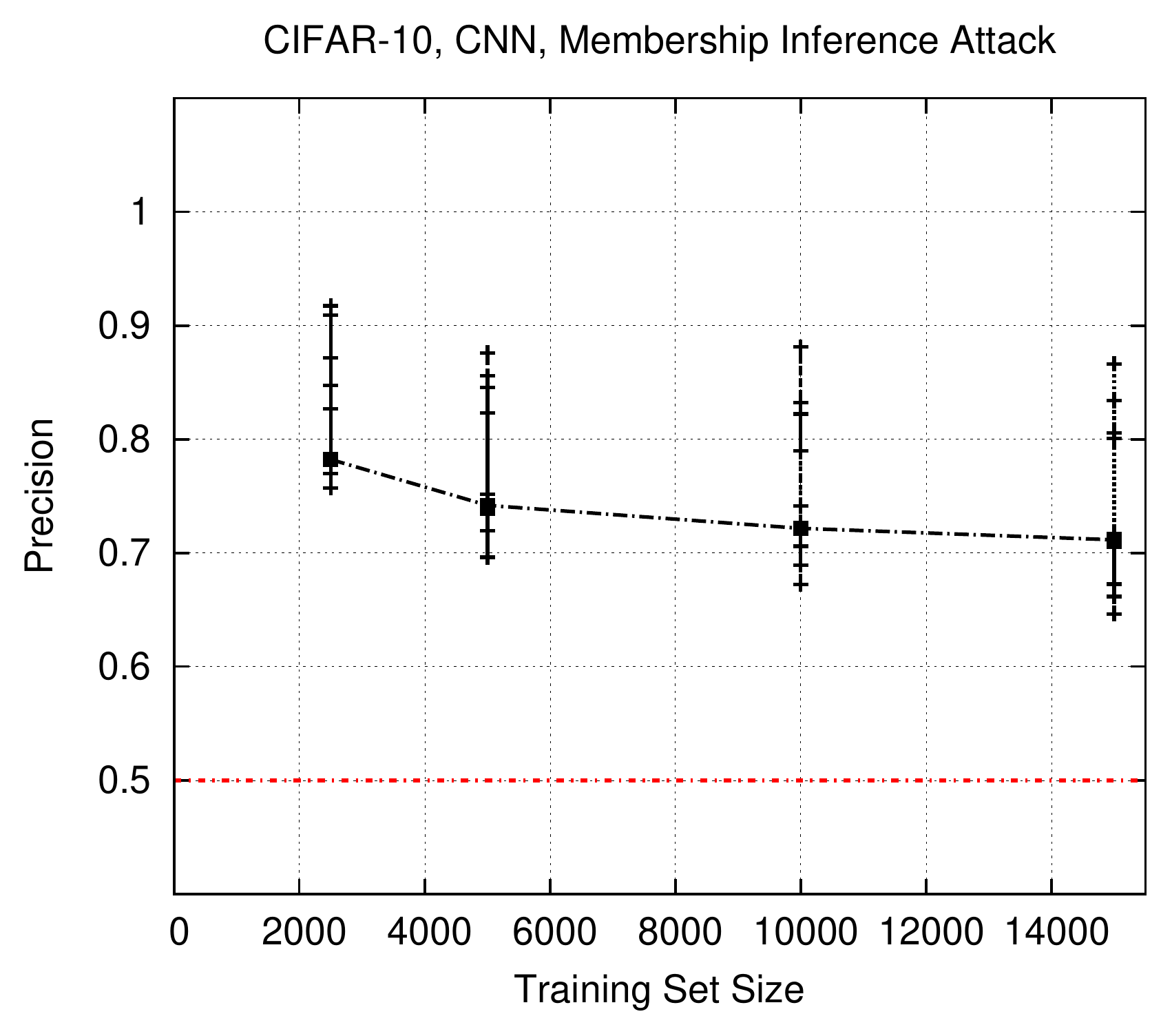}
\includegraphics[width=0.67\columnwidth]{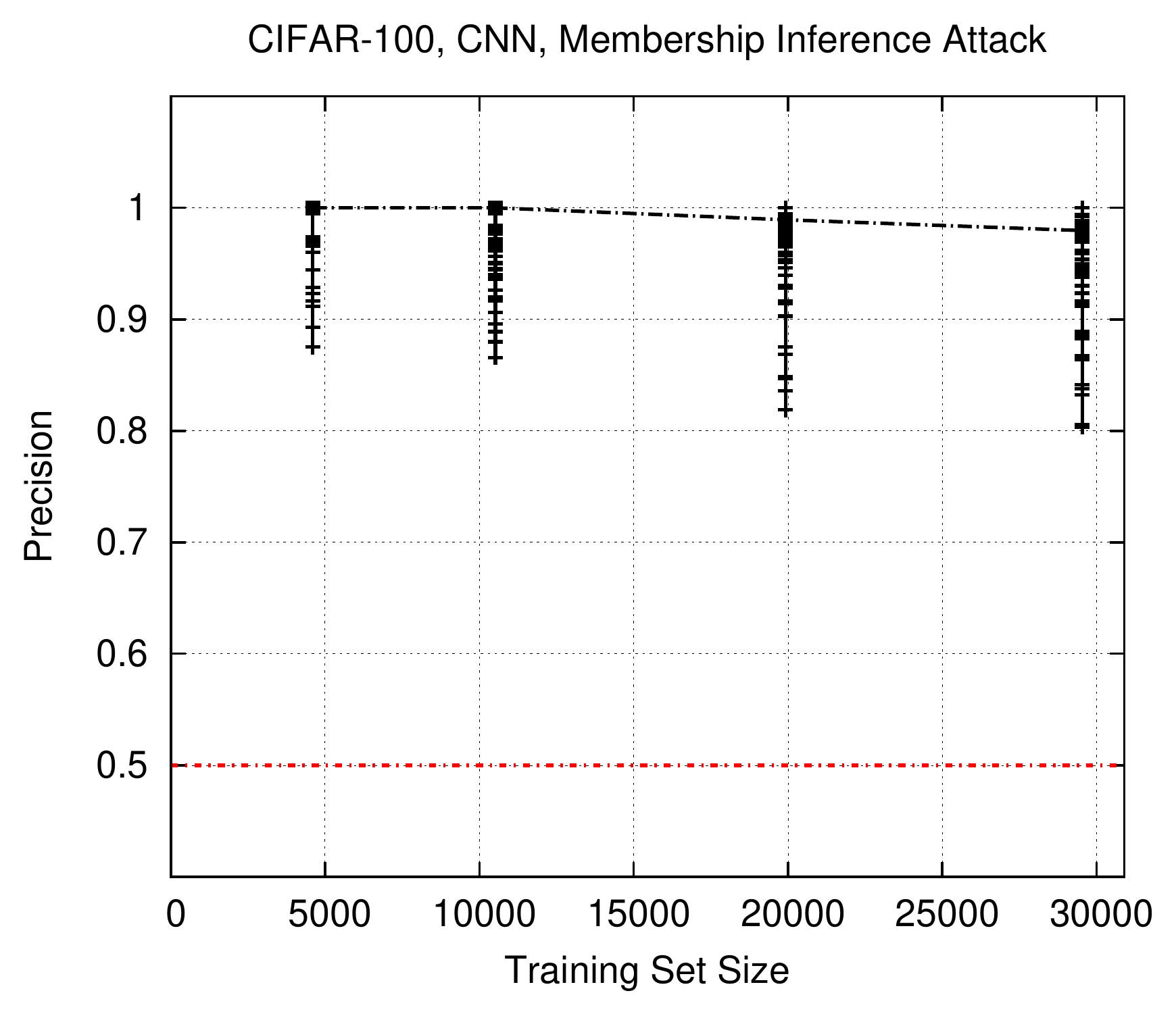}
\caption{Precision of the membership inference attack against
neural networks trained on CIFAR datasets.  The graphs show precision
for different classes while varying the size of the training datasets.
The median values are connected across different training set sizes.
The median precision (from the smallest dataset size to largest) is $0.78,
0.74, 0.72, 0.71$ for CIFAR-10 and $1, 1, 0.98, 0.97$ for CIFAR-100.
Recall is almost $1$ for both datasets.  The figure on the left
shows the per-class precision (for CIFAR-10).  Random guessing accuracy
is $0.5$.}\label{fig:cifar}

\end{figure*}

\begin{figure*}[t]
\centering
\includegraphics[width=0.69\columnwidth]{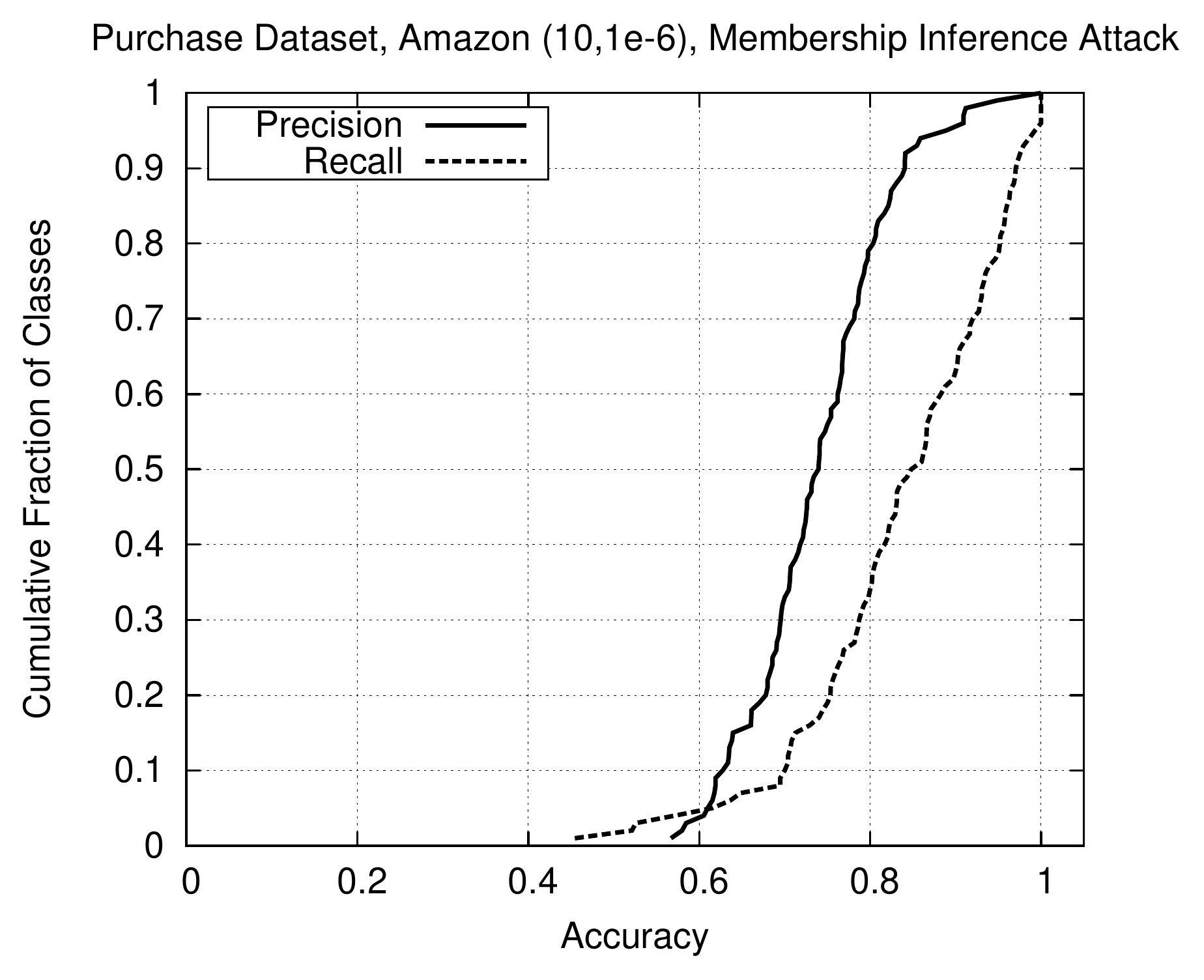}
\includegraphics[width=0.69\columnwidth]{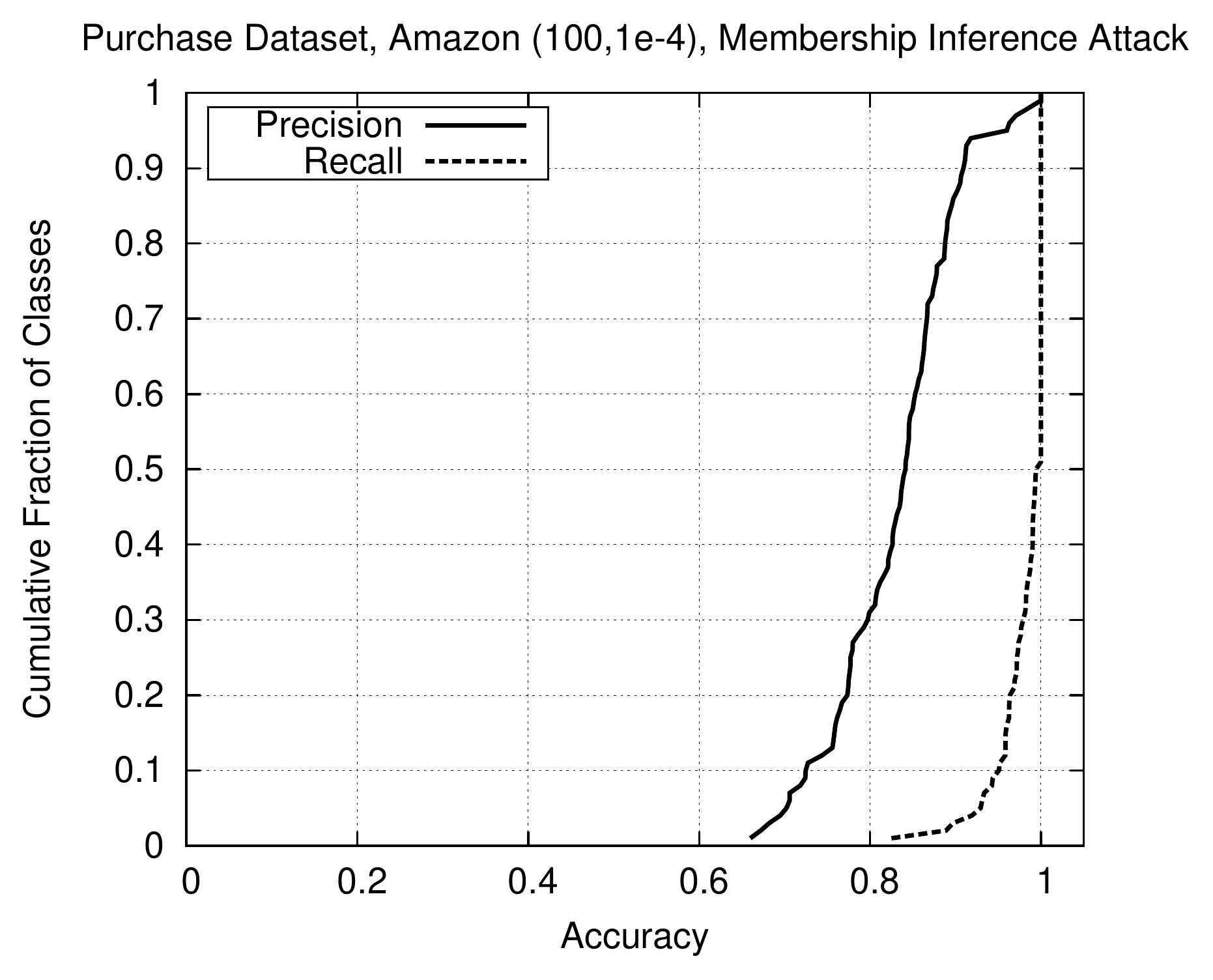}
\includegraphics[width=0.64\columnwidth]{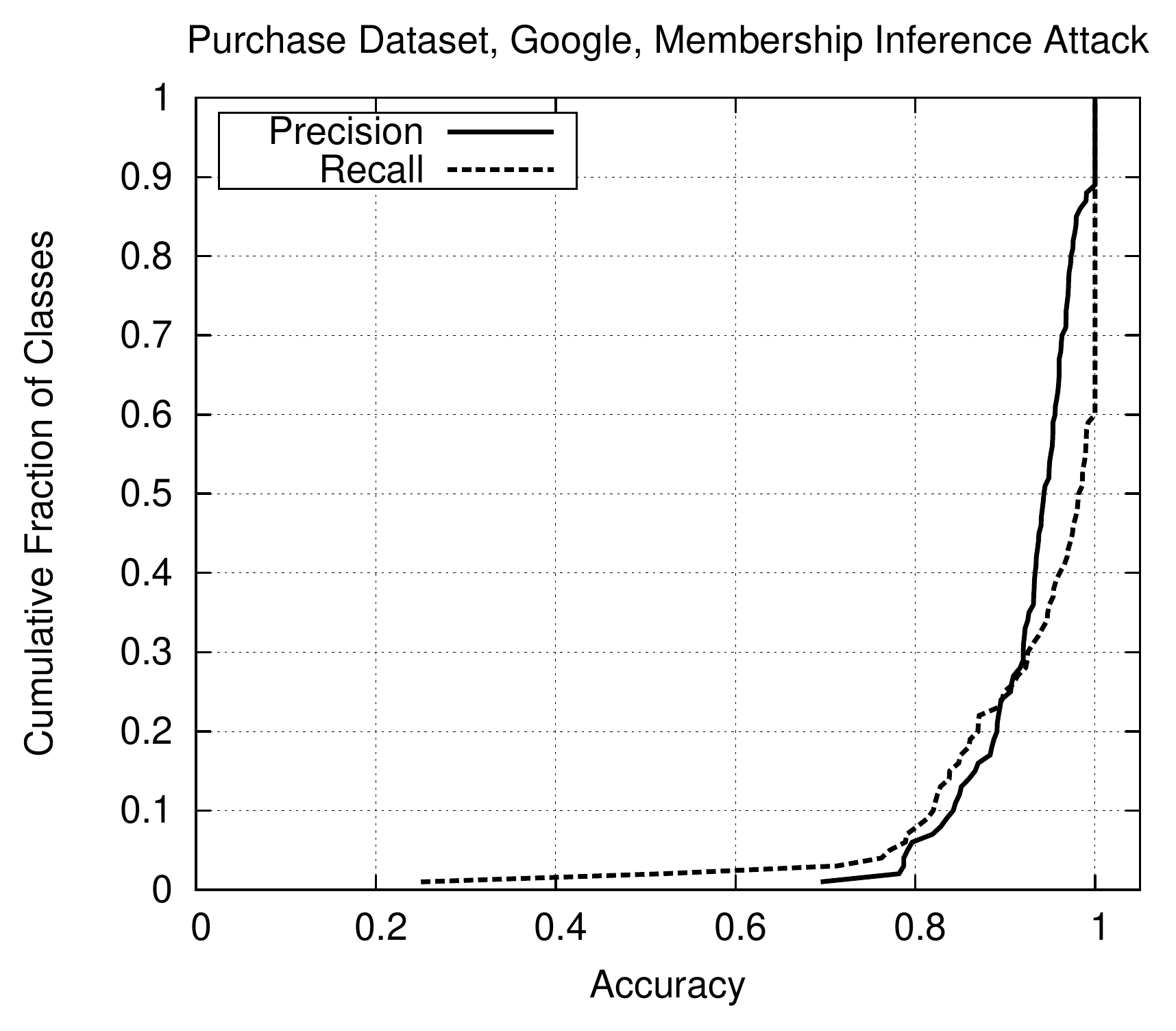}
\caption{Empirical CDF of the precision and recall of the membership
inference attack against different classes of the models trained using
Amazon ML (in two different configurations) and Google Prediction API
on $10,000$ purchase records.  $50, 75, 90$-percentile of precision is
$0.74, 0.79, 0.84$ on Amazon $(10,1e-6)$, $0.84, 0.88, 0.91$ on Amazon
$(100,1e-4)$, and $0.94, 0.97, 1$ on Google, respectively.  Recall is
close to 1.}\label{fig:googleamazon_purchase_membership}
\end{figure*}

\subsection{Target models}
\label{models}

We evaluated our inference attacks on three types of target models: two
constructed by cloud-based ``machine learning as a service'' platforms
and one we implemented locally.  In all cases, our attacks treat the
models as black boxes.  For the cloud services, we do not know the
type or structure of the models they create, nor the values of the
hyper-parameters used during the training process.

\paragraphbe{Machine learning as a service.}
The first cloud-based machine learning service in our study is Google
Prediction API.  With this service, the user uploads a dataset and obtains
an API for querying the resulting model.  There are no configuration
parameters that can be changed by the user.

The other cloud service is Amazon ML.  The user cannot choose the type
of the model but can control a few meta-parameters.  In our experiments,
we varied the \emph{maximum number of passes} over the training data
and L2 \emph{regularization amount}.  The former determines the number
of training epochs and controls the convergence of model training;
its default value is $10$.  The latter tunes how much regularization
is performed on the model parameters in order to avoid overfitting.
We used the platform in two configurations: the default setting ($10$,
$1e-6$) and ($100$, $1e-4$).

\paragraphbe{Neural networks.}
Neural networks have become a very popular approach to
large-scale machine learning.  We use Torch7 and its nn
packages,\footnote{\url{https://github.com/torch/nn}} a deep-learning
library that has been used and extended by major Internet companies such
as Facebook.\footnote{\url{https://github.com/facebook/fblualib}}

On CIFAR datasets, we train a standard convolutional neural network (CNN)
with two convolution and max pooling layers plus a fully connected layer
of size $128$ and a $\mathtt{SoftMax}$ layer.  We use $\mathtt{Tanh}$
as the activation function.  We set the learning rate to $0.001$,
the learning rate decay to $1e-07$, and the maximum epochs of training
to $100$.

On the purchase dataset (see Section~\ref{sec:data}), we train a fully
connected neural network with one hidden layer of size $128$ and a SoftMax
layer.  We use $\mathtt{Tanh}$ as the activation function.
We set the learning rate to $0.001$, the learning rate decay to $1e-07$,
and the maximum epochs of training to $200$.

\subsection{Experimental setup}

The training set and the test set of each target and shadow model are
randomly selected from the respective datasets, have the same size, and
are disjoint.  There is no overlap between the datasets of the target
model and those of the shadow models, but the datasets used for different
shadow models can overlap with each other.

We set the training set size to $10,000$ for the purchase dataset as well
as the Texas hospital-stay dataset, Adult dataset and the MNIST dataset.
We set it to $1,200$ for the location dataset.  We vary the size of the
training set for the CIFAR datasets, to measure the difference in the
attack accuracy.  For the CIFAR-10 dataset, we choose $2,500$; $5,000$;
$10,000$; and $15,000$.  For the CIFAR-100 dataset, we choose $4,600$;
$10,520$; $19,920$; and $29,540$.

The experiments on the CIFAR datasets were run locally, against our own
models, so we can vary the model's configuration and measure the impact
on the attack accuracy.  The experiments on the other datasets (purchases
with $\{2, 10, 20, 50, 100\}$ classes, Texas hospital stays, locations,
Adult, and MNIST) were run against models trained using either Google or
Amazon services, where we have no visibility into their choice of the
model type and structure and little control over the training process
(see Section~\ref{models}).

For the purchase dataset, we built target models on all platforms
(Google, Amazon, local neural networks) employing the same training
dataset, thus enabling us to compare the leakage from different models.
We used similar training architectures for the attack models across
different platforms: either a fully connected neural network with one
hidden layer of size 64 with ReLU (rectifier linear units) activation
functions and a SoftMax layer, or a Google-trained black-box model.

We set the number of shadow models to $100$ for the CIFAR datasets, $20$
for the purchase dataset, $10$ for the Texas hospital-stay dataset,
$60$ for the location dataset, $50$ for the MNIST dataset, and $20$
for the Adult dataset.  Increasing the number of shadow models would
increase the accuracy of the attack but also its cost.

\begin{figure}[t!]
\centering
\includegraphics[width=0.78\columnwidth]{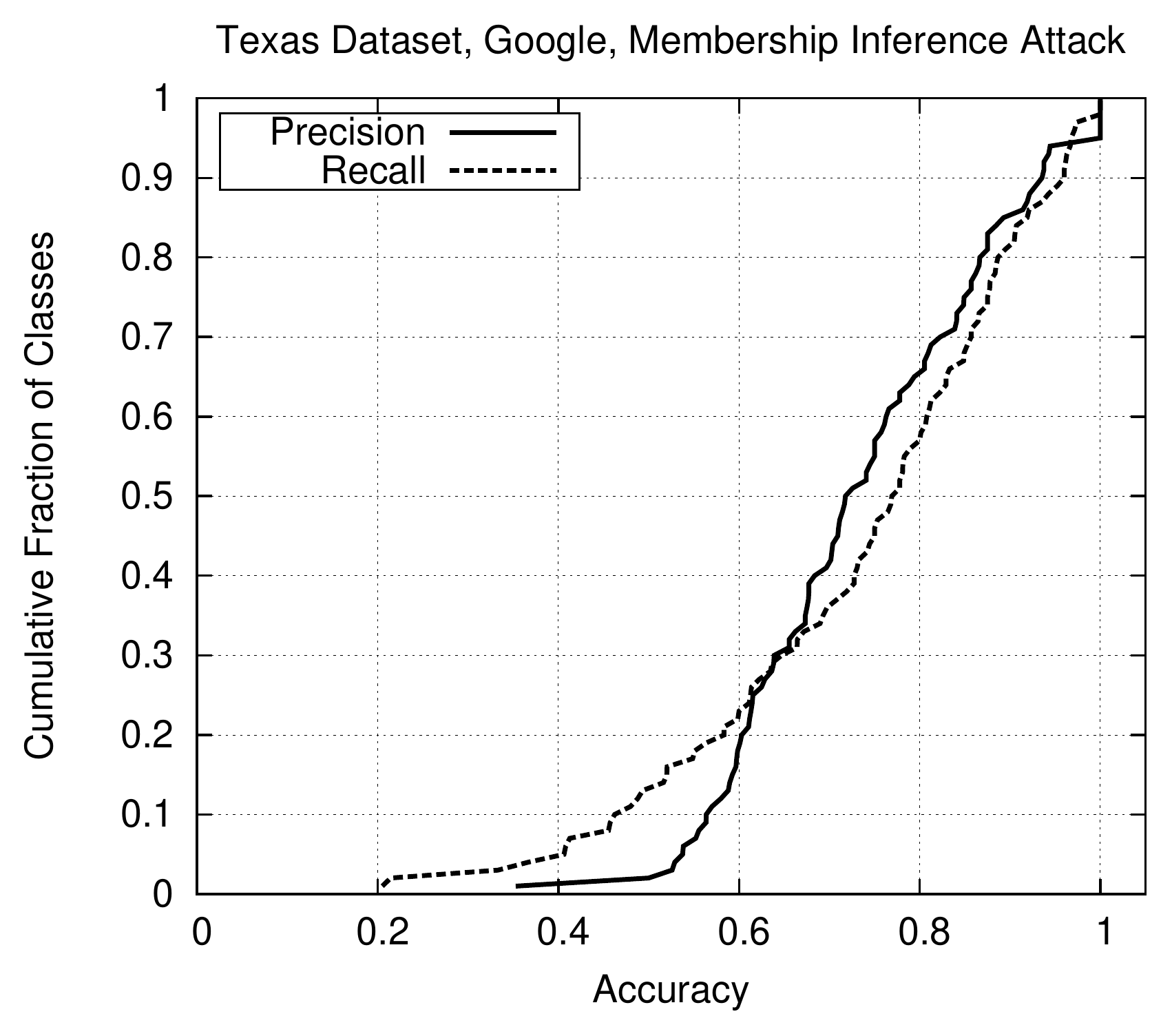}
\caption{Precision and recall of the membership inference attack against
the classification model trained using Google Prediction API on the
Texas hospital-stay dataset.}
\label{fig:google_texas_membership}
\end{figure}

\subsection{Accuracy of the attack}
\label{results}

The attacker's goal is to determine whether a given record was part
of the target model's training dataset.  We evaluate this attack by
executing it on randomly reshuffled records from the target's training and
test datasets.  In our attack evaluation, we use sets of the same size
(i.e, equal number of members and non-members) in order to maximize the
uncertainty of inference, thus the baseline accuracy is $0.5$.

We evaluate the attack using the standard \emph{precision} and
\emph{recall} metrics.  Precision is the fraction of the records
inferred as members of the training dataset that are indeed members.
Recall measures coverage of the attack, i.e., the fraction of the
training records that the attacker can correctly infer as members.
Most measurements are reported per class because the accuracy of the
attack can vary considerably for different classes.  This is due to the
difference in size and composition of the training data belonging to
each class and highly depends on the dataset.

The test accuracy of our target neural-network models with the largest
training datasets ($15,000$ and $29,540$ records, respectively) is $0.6$
and $0.2$ for CIFAR-10 and CIFAR-100, respectively.  The accuracy is low,
indicating that the models are heavily overfitted on their training sets.
Figure~\ref{fig:cifar} shows the results of the membership inference
attack against the CIFAR models.  For both CIFAR-10 and CIFAR-100, the
attack performs much better than the baseline, with CIFAR-100 especially
vulnerable.

\begin{figure}[t!]
\centering
\includegraphics[width=0.78\columnwidth]{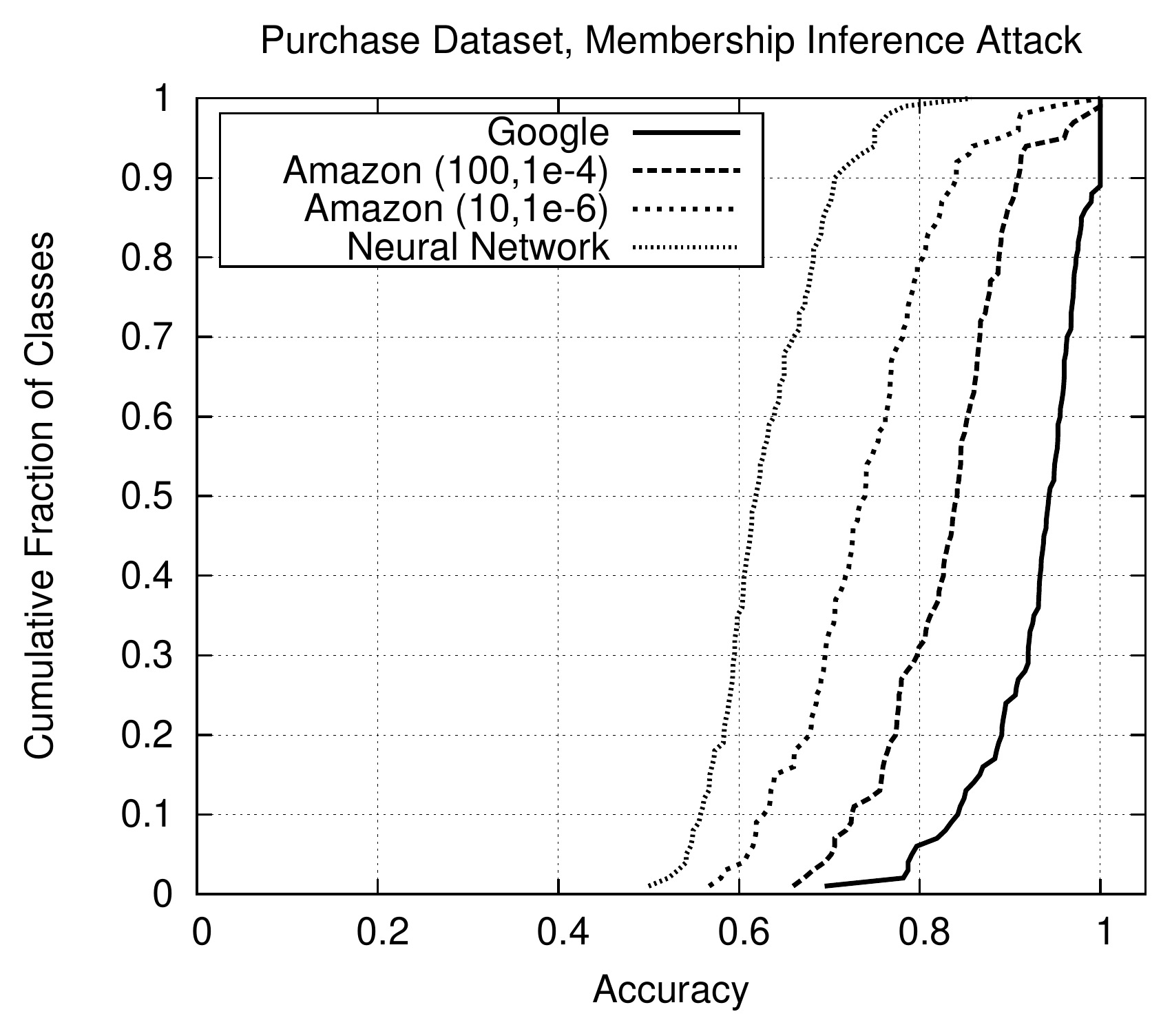}
\caption{Precision of the membership inference attack against models
trained on the same datasets but using different platforms.  The attack
model is a neural network.}
\label{fig:allplatforms_purchase_membership}
\end{figure}

Table~\ref{table:target-accuracy} shows the training and test accuracy
of the models constructed using different machine learning platforms
for the purchase dataset with 100 classes.  Large gaps between training
and test accuracy indicate overfitting.  Larger test accuracy indicates
better generalizability and higher predictive power.

Figure~\ref{fig:googleamazon_purchase_membership} shows the
results of the membership inference attack against the black-box
models trained by Google's and Amazon's machine learning platforms.
Figure~\ref{fig:allplatforms_purchase_membership} compares precision of
the attacks against these models with the attacks against a neural-network
model trained on the same data.  Models trained using Google Prediction
API exhibit the biggest leakage.

For the Texas hospital-stay dataset, we evaluated our attack against a
Google-trained model.  The training accuracy of the target model is $0.66$
and its test accuracy is $0.51$.  Figure~\ref{fig:google_texas_membership}
shows the accuracy of membership inference.  Precision is mostly above
$0.6$, and for half of the classes, it is above $0.7$.  Precision is
above $0.85$ for more than $20$ classes.

\begin{figure}[t!]
\centering
\includegraphics[width=0.78\columnwidth]{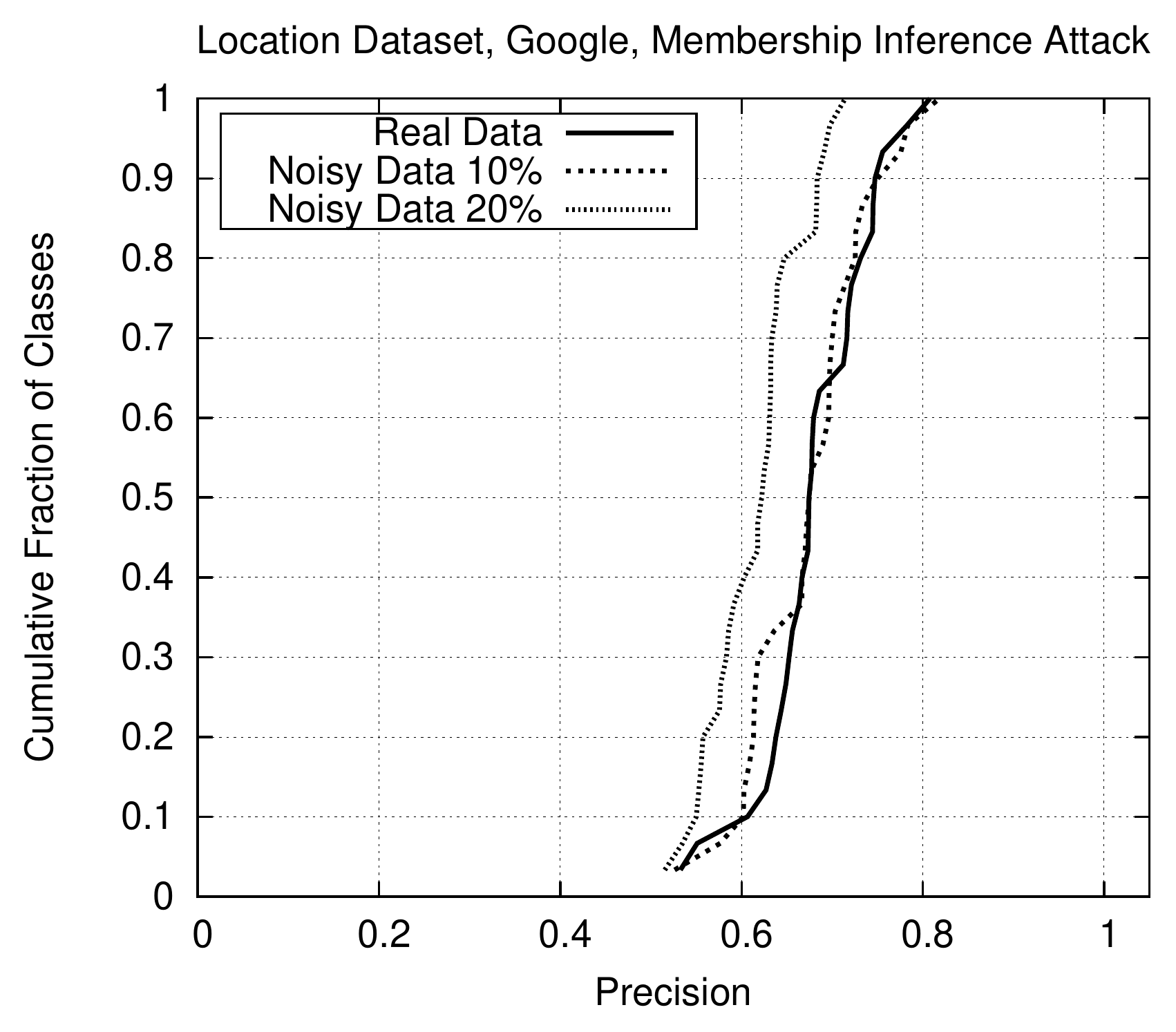}
\caption{Empirical CDF of the precision of the membership inference
attack against the Google-trained model for the location dataset.
Results are shown for the shadow models trained on real data and for
the shadow models trained on noisy data with $10\%$ and $20\%$ noise
(i.e., $x\%$ of features are replaced with random values).  Precision of
the attack over all classes is $0.678$ (real data), $0.666$ (data with
$10\%$ noise), and $0.613$ (data with $20\%$ noise).  The corresponding
recall of the attack is $0.98$, $0.99$, and $1.00$, respectively.}
\label{fig:google_location_membership}
\end{figure}

For the location dataset, we evaluated our attacks
against a Google-trained model.  The training accuracy of
the target model is $1$ and its test accuracy is $0.66$.
Figure~\ref{fig:google_location_membership} shows the accuracy of
membership inference.  Precision is between $0.6$ and $0.8$, with an
almost constant recall of $1$.

\begin{table}
\begin{center}
\begin{tabular}{l | r r }
  {\em ML Platform} & {\em Training}    & {\em Test}     \\\hline
  Google            & 0.999   & 0.656   \\
  Amazon (10,1e-6) \qquad\qquad  & 0.941   & 0.468   \\
  Amazon (100,1e-4) & 1.00      & 0.504    \\
  Neural network    & 0.830   & 0.670\\
\end{tabular}
\end{center}
\caption{Training and test accuracy of the models constructed using
different ML-as-a-service platforms on the purchase dataset (with 100
classes).}
\label{table:target-accuracy} \end{table}

\subsection{Effect of the shadow training data}
\label{sec:eval:noisysynthetic}

Figure~\ref{fig:google_location_membership} reports precision of the
attacks trained on the shadow models whose training datasets are noisy
versions of the real data (disjoint from the target model's training
dataset but sampled from the same population).  Precision drops as the
amount of noise increases, but the attack still outperforms the baseline
and, even with 10\% of the features in the shadows' training data replaced
by random values, matches the original attack.  This demonstrates that
\textbf{our attacks are robust even if the attacker's assumptions
about the distribution of the target model's training data are not
very accurate}.

\begin{figure}[t!]
\centering
\includegraphics[width=0.78\columnwidth]{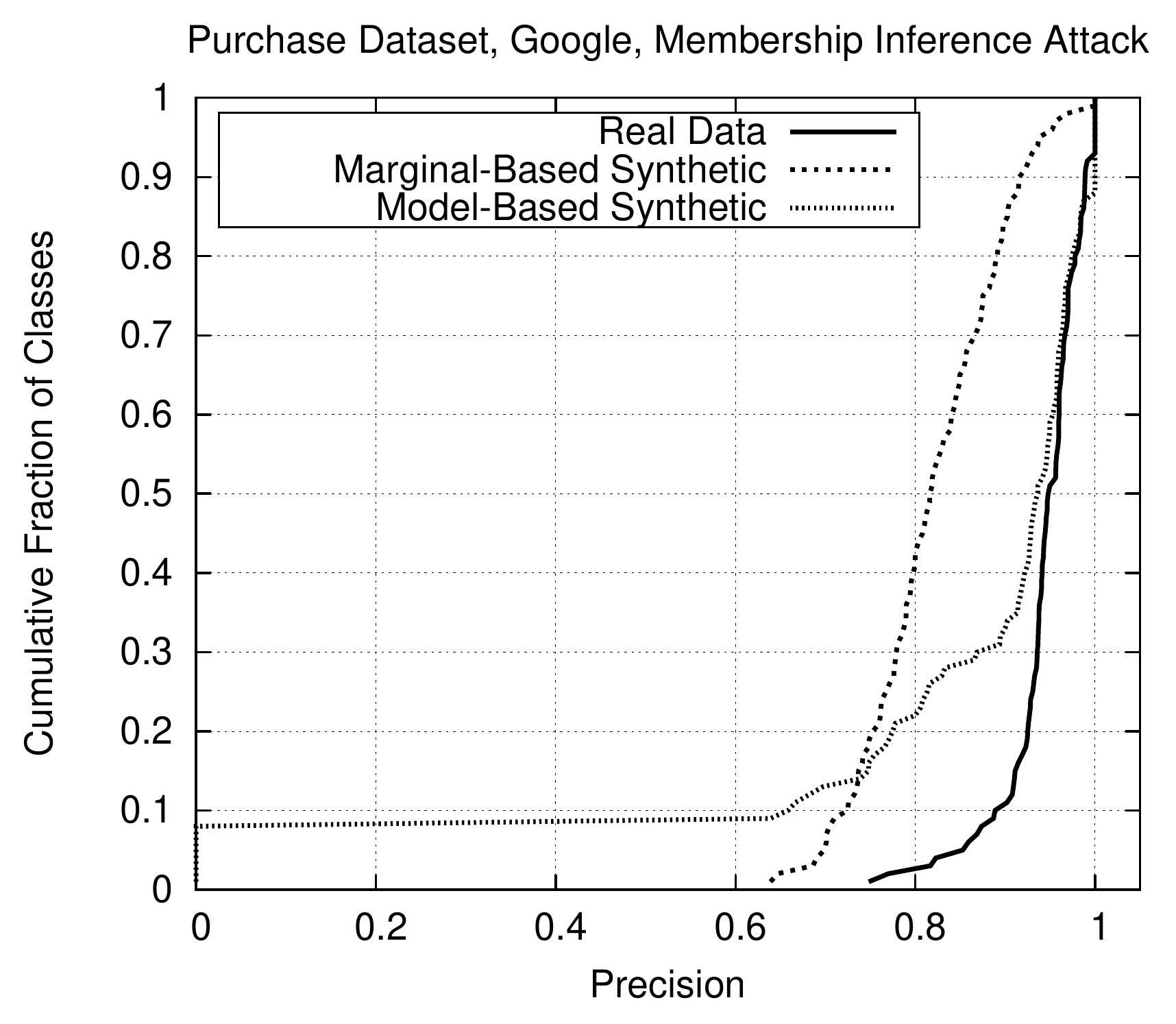}
\caption{Empirical CDF of the precision of the membership inference
attack against the Google-trained model for the purchase dataset.
Results are shown for different ways of generating training data for
the shadow models (real, synthetic generated from the target model,
synthetic generated from marginal statistics).  Precision of the attack
over all classes is $0.935$ (real data), $0.795$ (marginal-based synthetic
data), and $0.896$ (model-based synthetic data).  The corresponding
recall of the attack is $0.994$, $0.991$, and $0.526$, respectively.}
\label{fig:google_purchase_membership_synthetic}
\end{figure}

Figure~\ref{fig:google_purchase_membership_synthetic} reports precision
of the attacks when the attacker has no real data (not even noisy) for
training his shadow models.  Instead, we used the marginal distributions
of individual features to generate $187,300$ synthetic purchase records,
then trained $20$ shadow models on these records.

We also generated $30,000$ synthetic records using the model-based
approach presented in Algorithm 1.  In our experiments with the purchase
dataset where records have $600$ binary features, we initialize $k$ to
$k_{max} = 128$ and divide it by $2$ when $rej_{max} = 10$ subsequent
proposals are rejected.  We set its minimum value $k_{min} = 4$.  In the
sampling phase, we set the minimum confidence threshold $conf_{min}$
to $0.2$.

For our final set of sampled records, the target model's confidence
in classifying the records is $0.24$ on average (just a bit over our
threshold $conf_{min}=0.2$).  On average, each synthetic record needed
$156$ queries (of proposed records) during our hill-climbing two-phase
process (see Section~\ref{shadowtrain}).  We trained $8$ shadow models
on this data.

Figure~\ref{fig:google_purchase_membership_synthetic} compares precision
of the attacks when shadow models are trained on real data versus shadow
models trained on synthetic data.  The overall precision is $0.935$
on real data compared to $0.795$ for marginal-based synthetics and
$0.895$ for model-based synthetics.  The accuracy of the attack using
marginal-based synthetic data is noticeably reduced versus real data, but
is nevertheless very high for most classes.  The attack using model-based
synthetic data exhibits dual behavior.  For most classes its precision
is high and close to the attacks that use real data for shadow training,
but for a few classes precision is very low (less than $0.1$).

The reason for the attack's low precision on some classes is that the
target classifier cannot confidently model the distribution of data
records belonging to these classes\textemdash because it has not seen
enough examples.  These classes are under-represented in the target
model's training dataset.  For example, each of the classes where the
attack has less than $0.1$ precision contributes under $0.6\%$ of the
target model's training dataset.  Some of these classes have fewer than
$30$ training records (out of $10,000$).  This makes it very difficult
for our algorithm to synthesize representatives of these classes when
searching the high-dimensional space of possible records.

For the majority of the target model's classes, our attack achieves
high precision.  This demonstrates that \textbf{a membership inference
attack can be trained with only black-box access to the target model,
without any prior knowledge about the distribution of the target model's
training data} if the attacker can efficiently generate inputs that are
classified by the target model with high confidence.

\begin{figure}
\centering
\includegraphics[width=0.78\columnwidth]{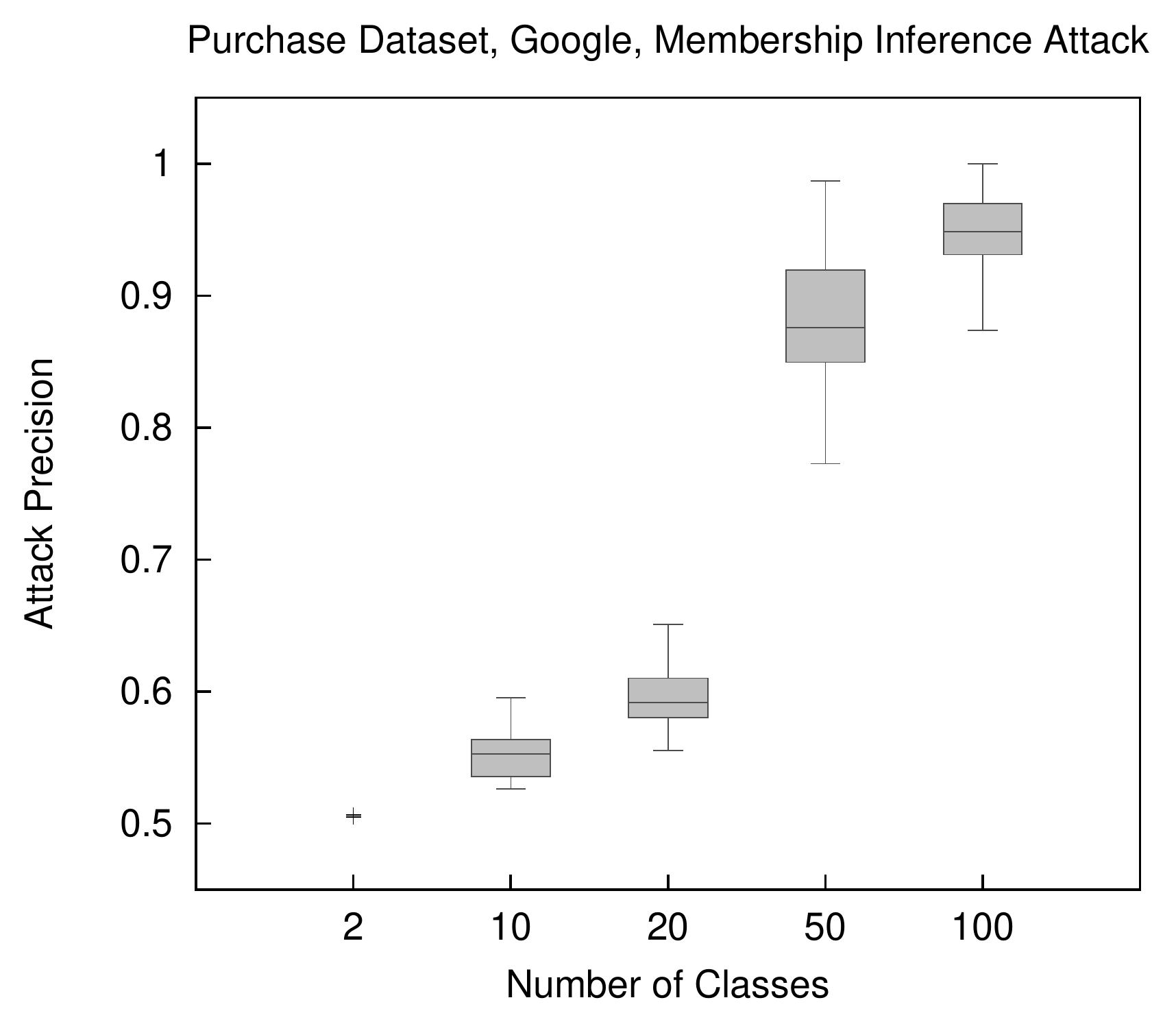}
\caption{Precision of the membership inference attack against different
purchase classification models trained on the Google platform.  The
boxplots show the distribution of precision over different classification
tasks (with a different number of classes).} \label{fig:purchase2_100}
\end{figure}

\begin{figure*}[t!]
\centering
\includegraphics[width=0.65\columnwidth]{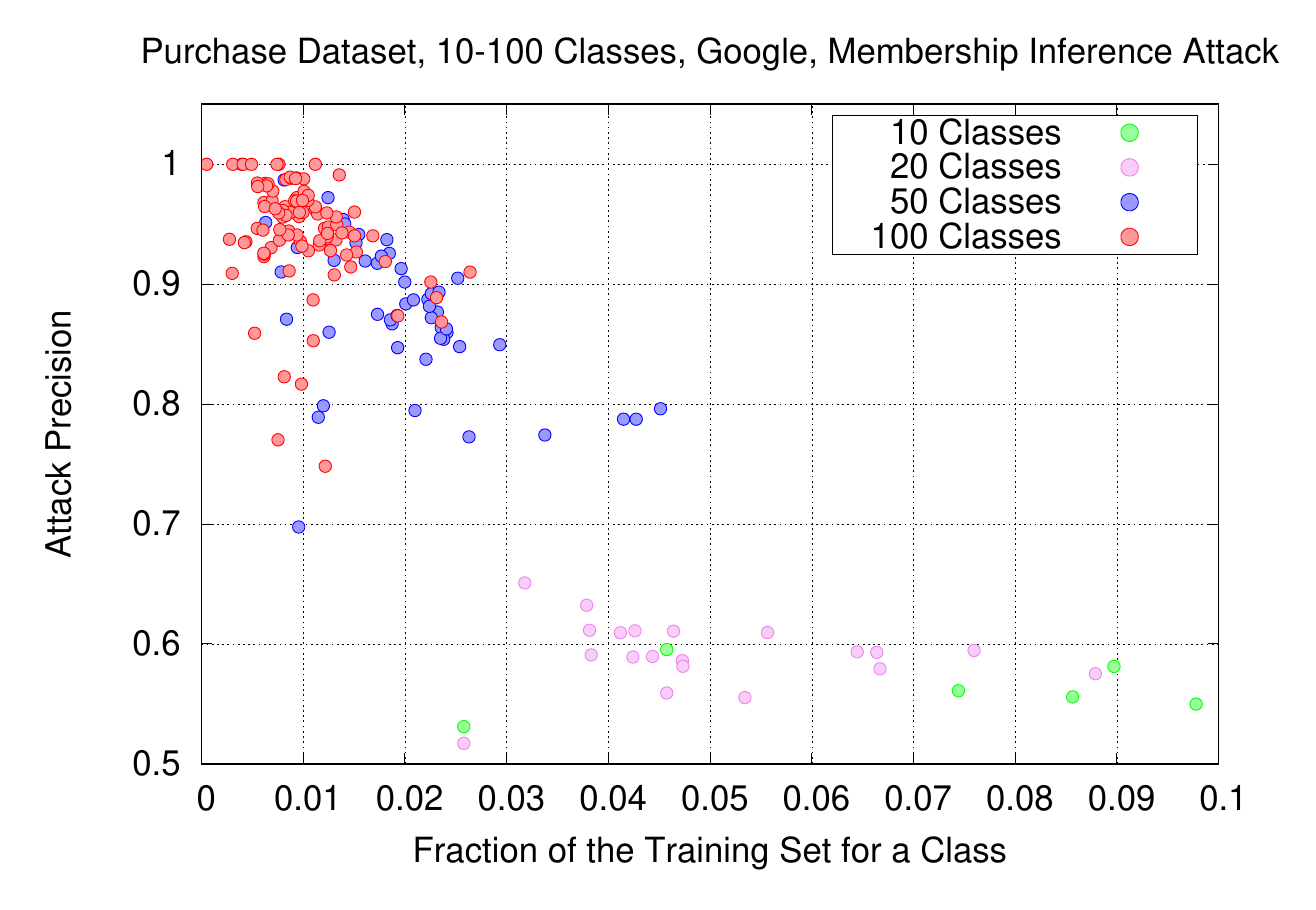}
\includegraphics[width=0.65\columnwidth]{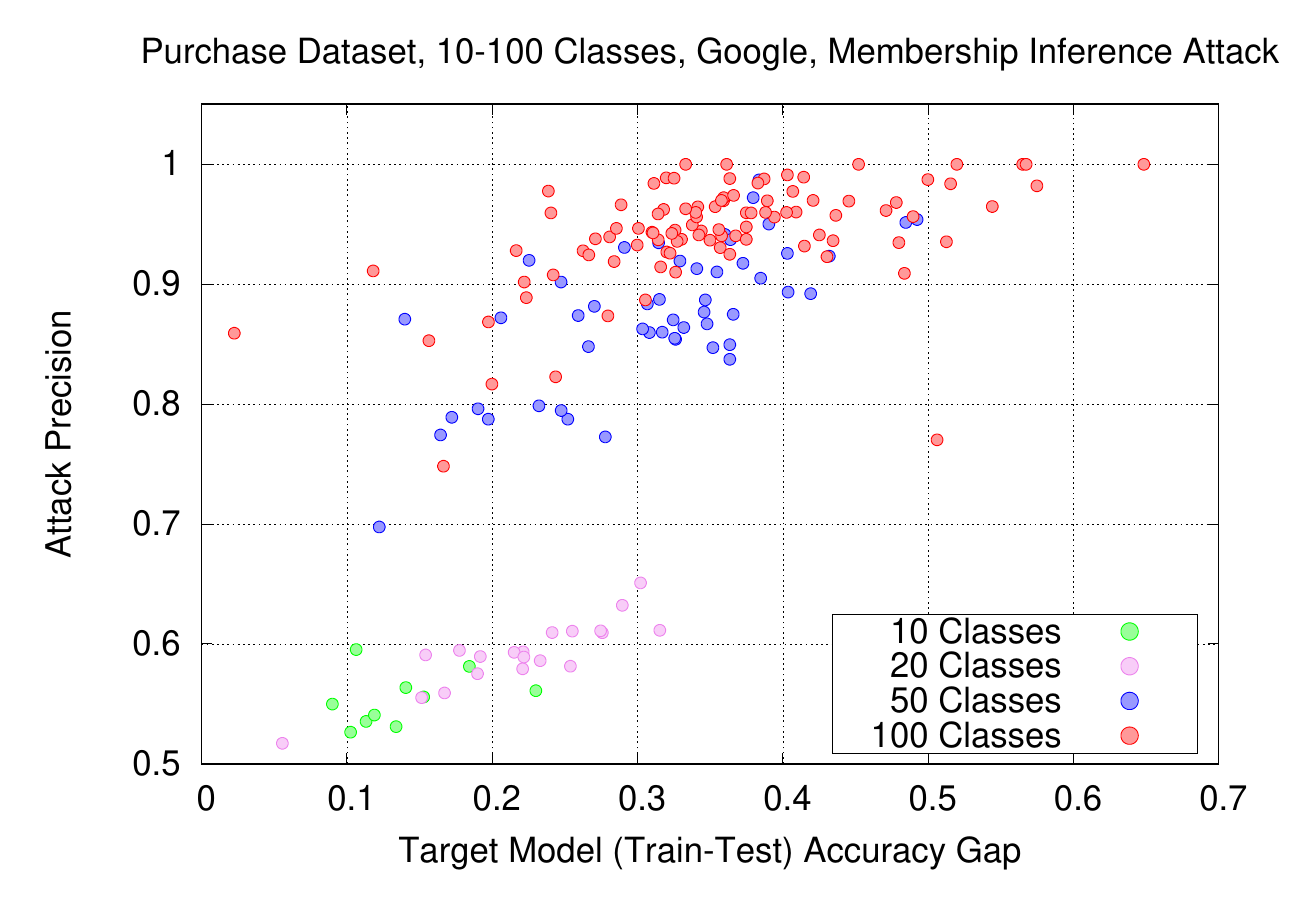}
\includegraphics[width=0.65\columnwidth]{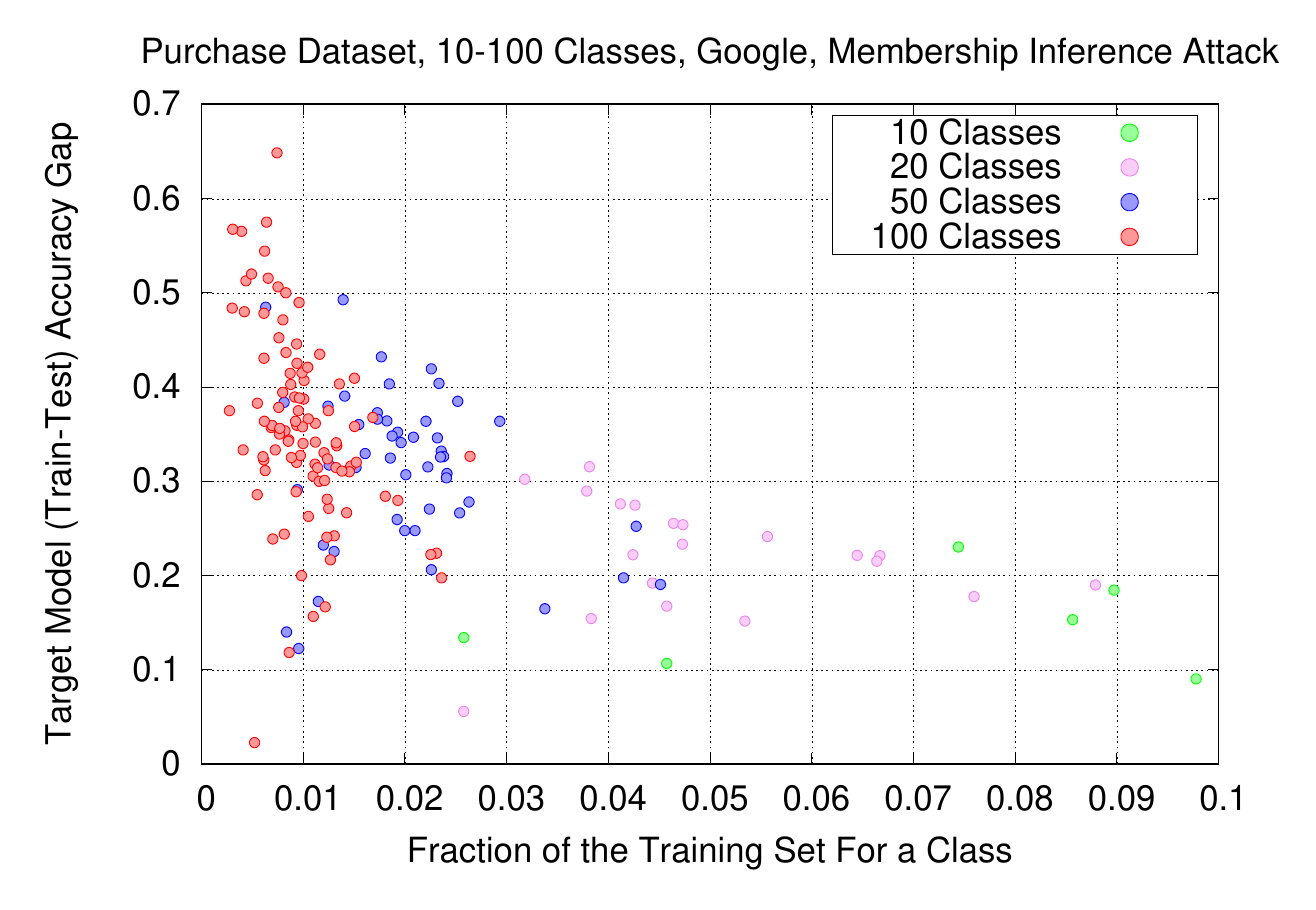}\\
\caption{Relationship between the precision of the membership inference
attack on a class and the (train-test) accuracy gap of the target model,
as well as the fraction of the training dataset that belongs to this
class.  Each point represent the values for one class.  The (train-test)
accuracy gap is a metric for generalization error~\cite{hardt2015train}
and an indicator of how overfitted the target model is.}
\label{fig:howitworks}
\end{figure*}

\subsection{Effect of the number of classes and training
data per class}

The number of output classes of the target model contributes to how much
the model leaks.  The more classes, the more signals about the internal
state of the model are available to the attacker.  This is one of the
reasons why the results in Fig.~\ref{fig:cifar} are better for CIFAR-100
than for CIFAR-10.  The CIFAR-100 model is also more overfitted to its
training dataset.  For the same number of training records per class,
the attack performs better against CIFAR-100 than against CIFAR-10.
For example, compare CIFAR-10 when the size of the training dataset is
$2,000$ with CIFAR-100 when the size of the training dataset is $20,000$.
The average number of data records per class is $200$ in both cases,
but the attack accuracy is much better (close to $1$) for CIFAR-100.

To quantify the effect that the number of classes has on the accuracy
of the attack, we trained target models using Google Prediction
API on the purchase dataset with $\{2,10,20,50,100\}$ classes.
Figure~\ref{fig:purchase2_100} shows the distribution of attack precision
for each model.  Models with fewer classes leak less information about
their training inputs.  As the number of classes increases, the model
needs to extract more distinctive features from the data to be able to
classify inputs with high accuracy.  Informally, models with more output
classes need to remember more about their training data, thus they leak
more information.

Figure~\ref{fig:howitworks} shows the relationship between the amount
of training data per class and the accuracy of membership inference.
This relationship is more complex, but, in general, the more data in the
training dataset is associated with a given class, the lower the attack
precision for that class.

Table~\ref{table:google} shows the precision of membership inference
against Google-trained models.  For the MNIST dataset, the training
accuracy of the target model is $0.984$ and its test accuracy is $0.928$.
The overall precision of the membership inference attack is $0.517$,
which is just slightly above random guessing.  The lack of randomness
in the training data for each class and the small number of classes
contribute to the failure of the attack.

For the Adult dataset, the training accuracy of the target model is
$0.848$ and its test accuracy is $0.842$.  The overall precision of the
attack is $0.503$, which is equivalent to random guessing.  There could
be two reasons for why membership inference fails against this model.
First, the model is not overfitted (its test and train accuracies are
almost the same).  Second, the model is a binary classifier, which
means that the attacker has to distinguish members from non-members by
observing the behavior of the model on essentially 1 signal, since the
two outputs are complements of each other.  This is not enough for our
attack to extract useful membership information from the model.

\begin{table}
\begin{center}
\begin{tabular}{l | c c c}
  {\em Dataset}		& {\em Training}	& {\em Testing}	& {\em Attack} \\
  					& {\em Accuracy}	& {\em Accuracy}& {\em Precision} \\\hline
  Adult				& 0.848				& 0.842			& 0.503 \\
  MNIST 			& 0.984				& 0.928			& 0.517 \\
  Location			& 1.000				& 0.673			& 0.678 \\
  Purchase (2)		& 0.999				& 0.984			& 0.505 \\
  Purchase (10)		& 0.999				& 0.866			& 0.550 \\
  Purchase (20)		& 1.000				& 0.781			& 0.590 \\
  Purchase (50)		& 1.000				& 0.693			& 0.860 \\
  Purchase (100)	& 0.999				& 0.659			& 0.935 \\
  TX hospital stays	& 0.668				& 0.517			& 0.657 \\
\end{tabular}
\end{center}
\caption{Accuracy of the Google-trained models and the corresponding
attack precision.}
\label{table:google} \end{table}

\begin{figure*}[t!]
\centering
\includegraphics[width=0.65\columnwidth]{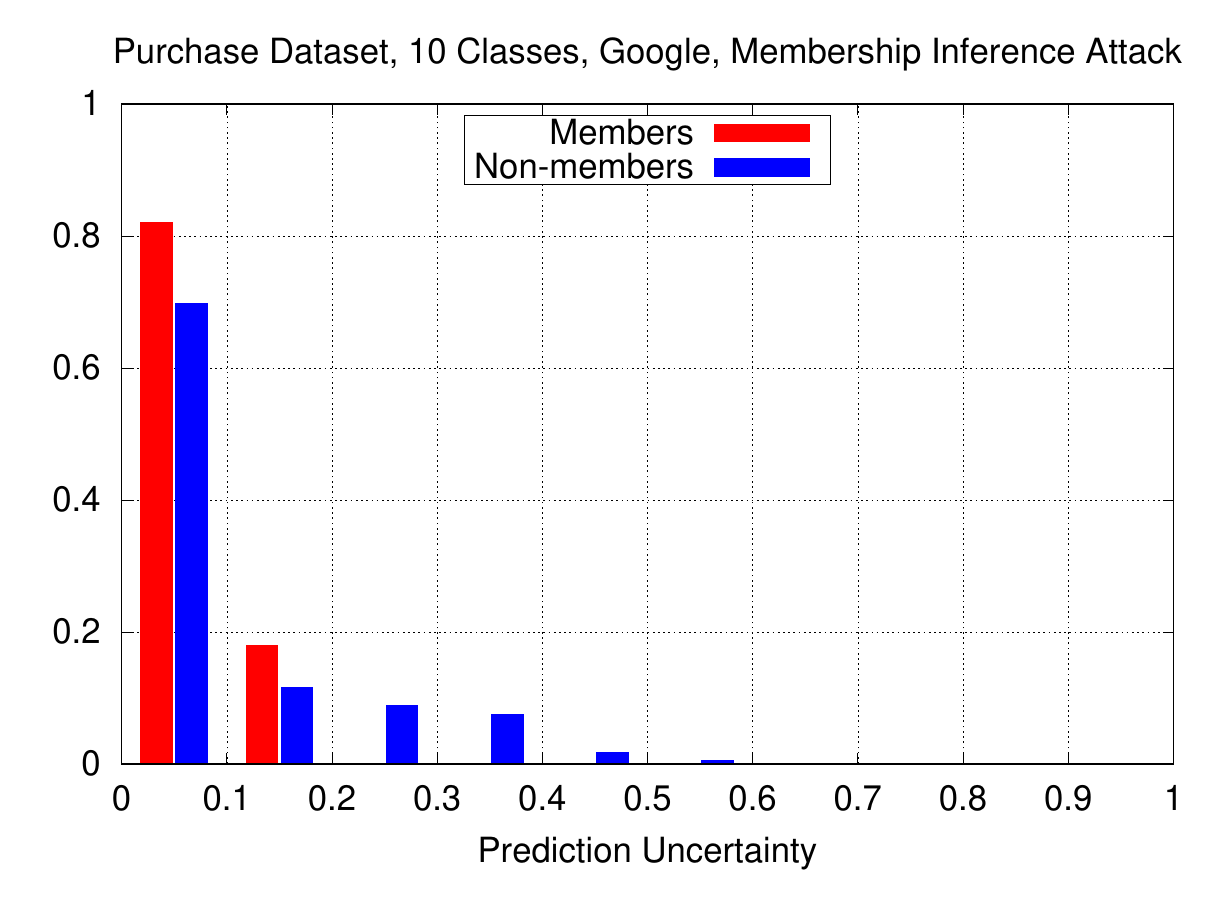}
\includegraphics[width=0.65\columnwidth]{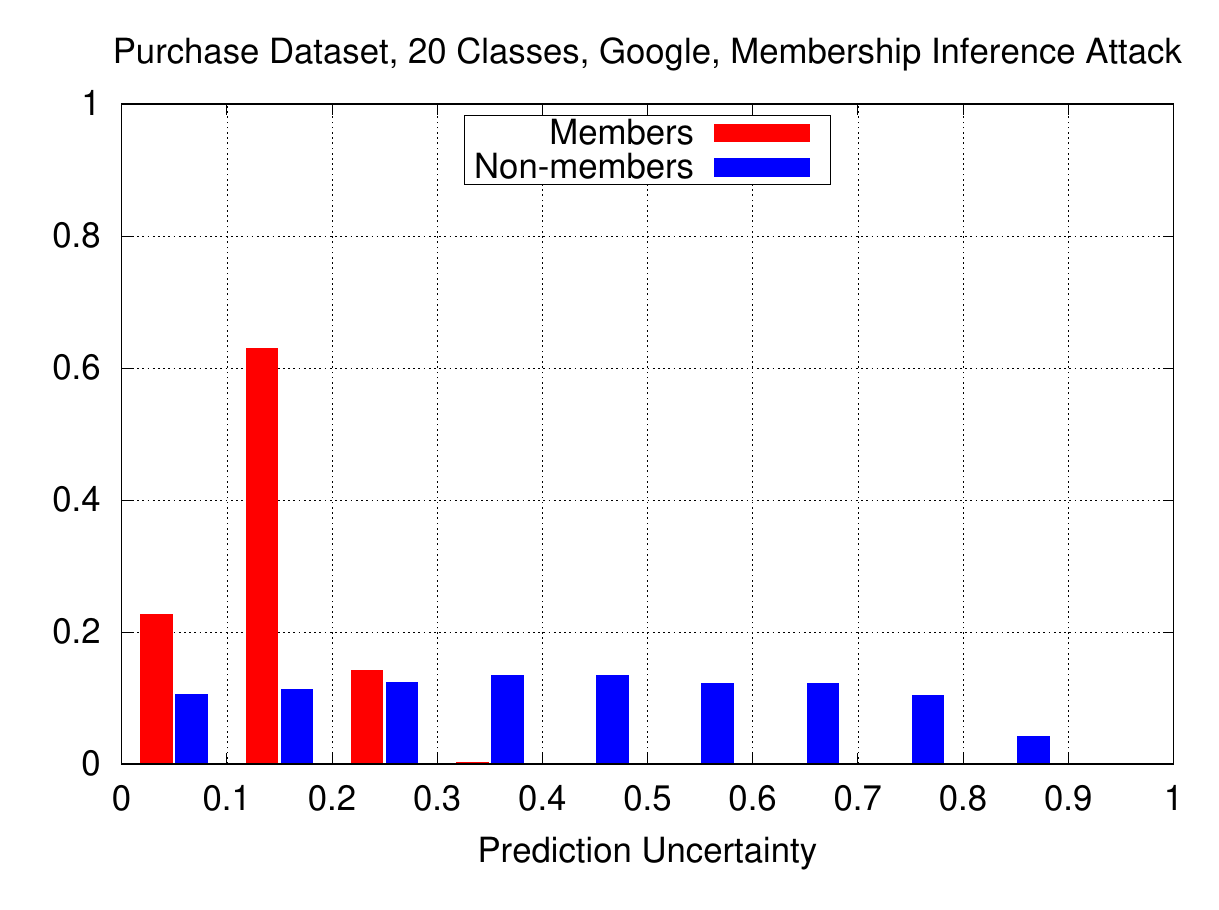}
\includegraphics[width=0.65\columnwidth]{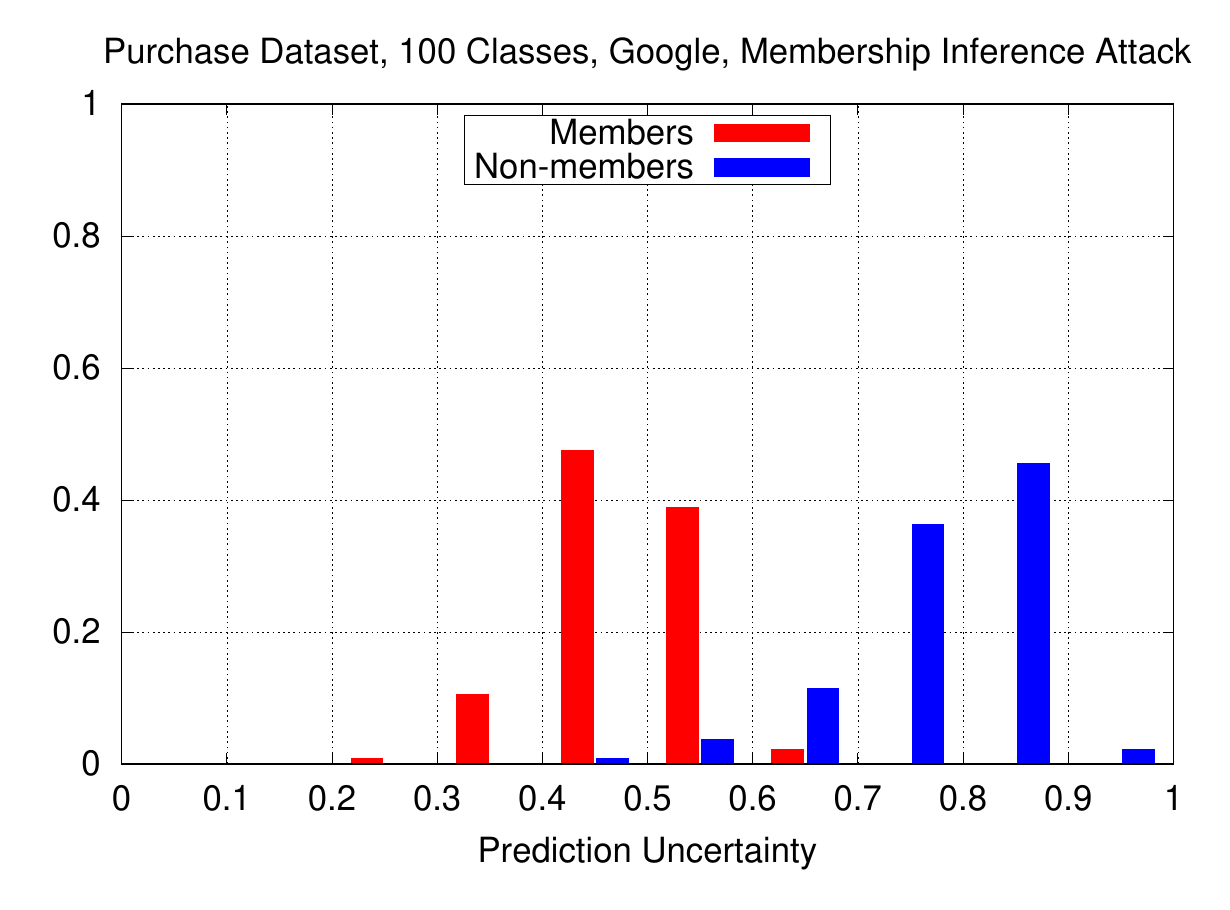}\\
\includegraphics[width=0.65\columnwidth]{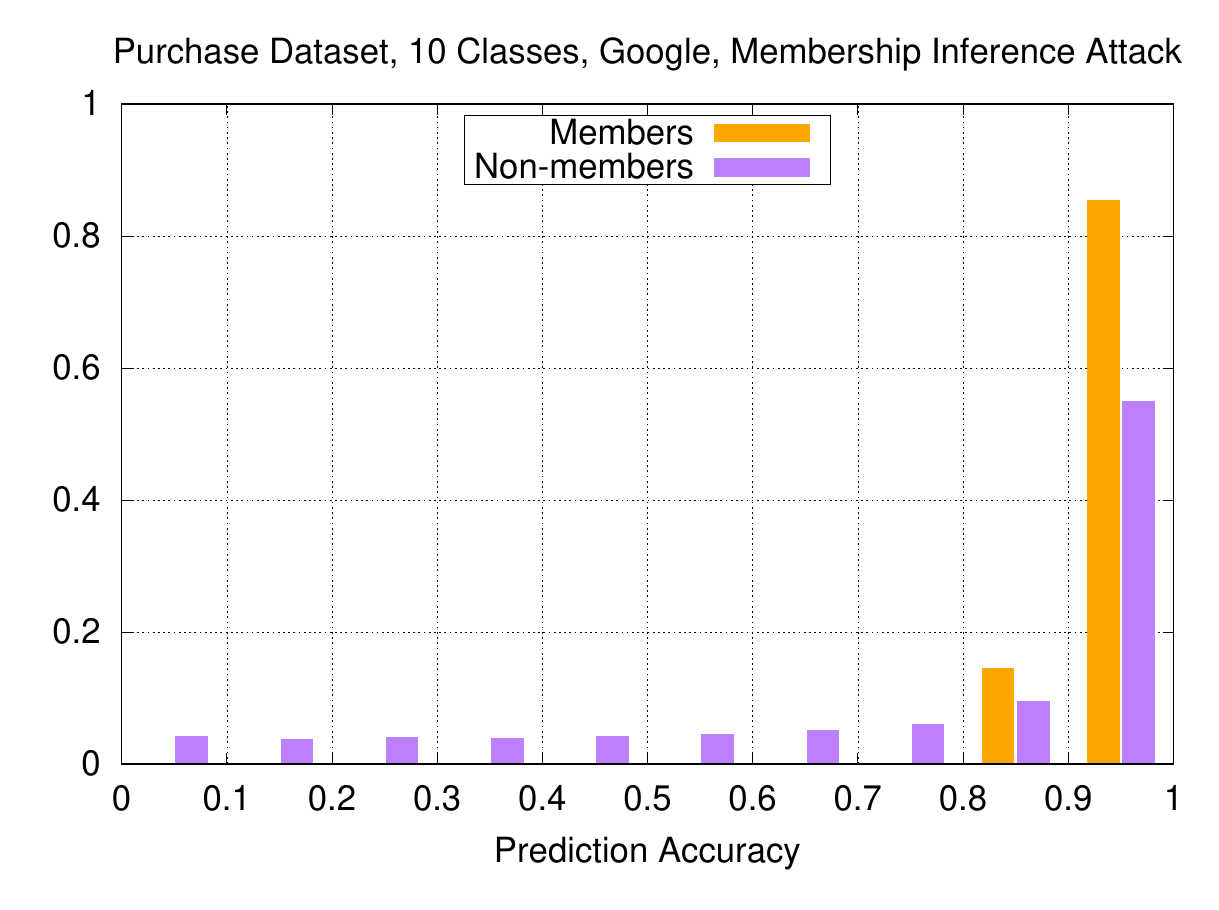} 
\includegraphics[width=0.65\columnwidth]{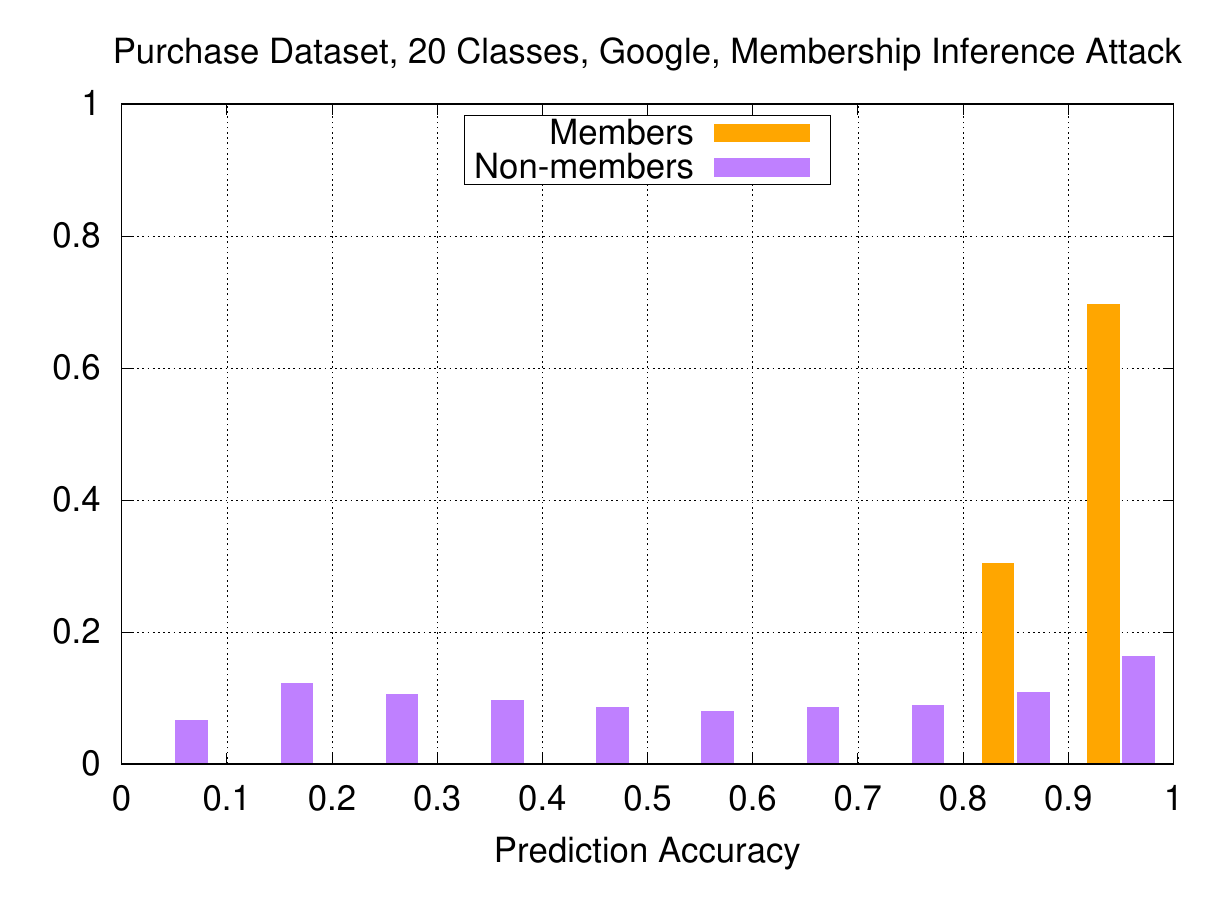} 
\includegraphics[width=0.65\columnwidth]{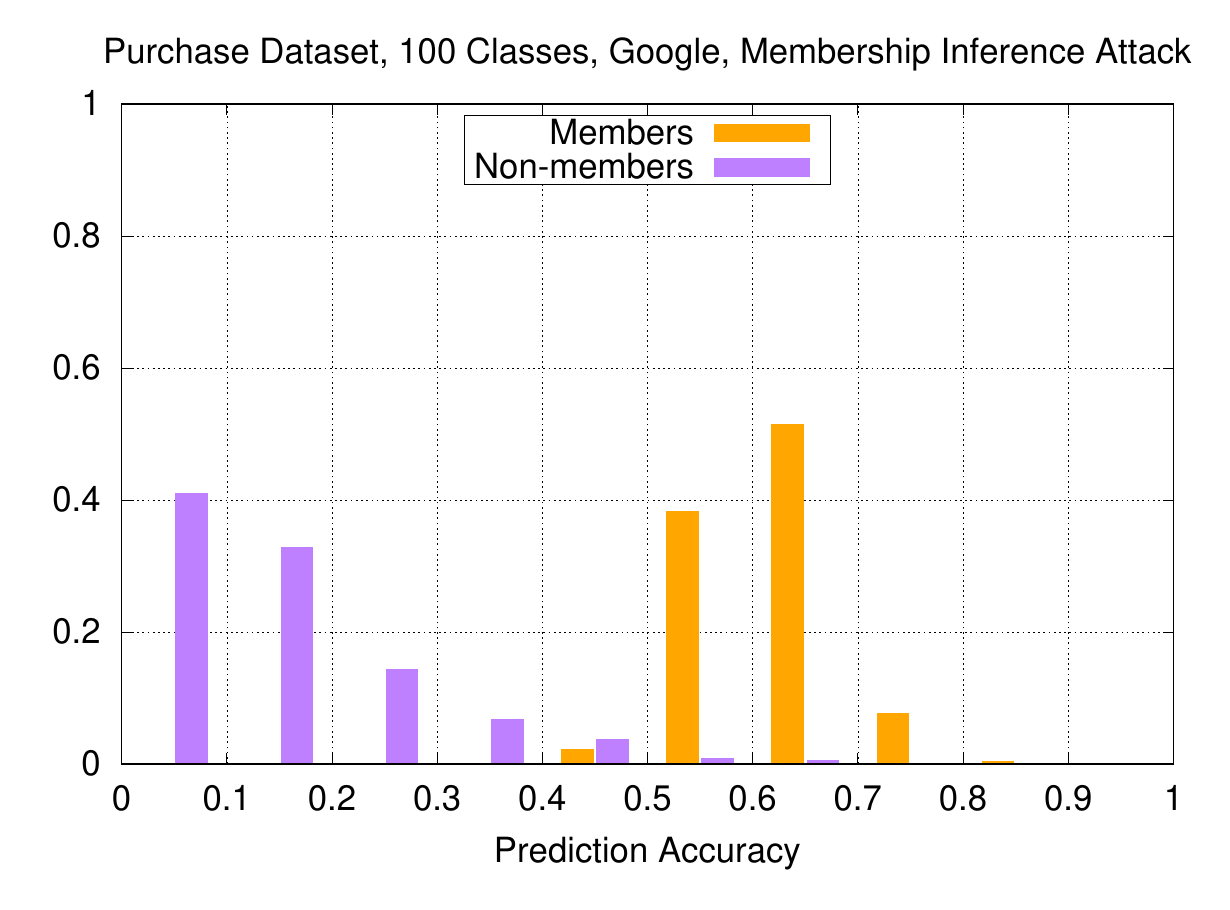}
\caption{Classification uncertainty (top row) and prediction accuracy
(bottom row) of the target model for the members of its training dataset
vs.\ non-members, visualized for several sample classes.  The difference
between the member and non-member output distributions is among the
factors that our attack exploits to infer membership.  The accuracy
of our attack is higher for the models where the two distributions are
more distinguishable
(See Table~\ref{table:google}).}
\label{fig:whyitworks}
\end{figure*}

\subsection{Effect of overfitting}

The more overfitted a model, the more it leaks\textemdash but only for
models of the same type.  For example, the Amazon-trained $(100,1e-4)$
model that, according to Table~\ref{table:target-accuracy}, is more
overfitted leaks more than the Amazon-trained $(10,1e-6)$ model.
However, they both leak less than the Google-trained model, even though
the Google model is less overfitted than one of the Amazon models and
has a much better predictive power (and thus generalizability) than
both Amazon models.  Therefore, \textbf{overfitting is not the only
factor that causes a model to be vulnerable to membership inference}.
The structure and type of the model also contribute to the problem.

In Figure~\ref{fig:howitworks}, we look deeper into the factors that
contribute to attack accuracy per class, including how overfitted
the model is and what fraction of the training data belongs to each
class.  The (train-test) accuracy gap is the difference between the
accuracy of the target model on its training and test data.  Similar
metrics are used in the literature to measure how overfitted a model
is~\cite{hardt2015train}.  We compute this metric for each class.  Bigger
gaps indicate that the model is overfitted on its training data for that
class.  The plots show that, as expected, bigger (train-test) accuracy
gaps are associated with higher precision of membership inference.

\section{Why Our Attacks Work}
\label{sec:whyattackworks}

Table~\ref{table:google} shows the relationship between the accuracy
of our membership inference attack and the (train-test) gap of the
target models.  Figure~\ref{fig:whyitworks} also illustrates how the
target models' outputs distinguish members of their training datasets
from the non-members.  This is the information that our attack exploits.

Specifically, we look at how accurately the model predicts the correct
label as well as its prediction uncertainty.  The accuracy for class $i$
is the probability that the model classifies an input with label $i$
as $i$.  Prediction uncertainty is the normalized entropy of the model's
prediction vector: $\frac{-1}{\log(n)} \sum_i p_i \log(p_i)$, where
$p_i$ is the probability that the input belongs to class $i$, and $n$
is the number of classes.  The plots show that there is an observable
difference between the output (both accuracy and uncertainty) of the
model on the member inputs versus the non-member inputs in the cases
where our attack is successful.

Success of membership inference is directly related to the (1)
generalizability of the target model and (2) diversity of its training
data.  If the model overfits and does not generalize well to inputs
beyond its training data, or if the training data is not representative,
the model leaks information about its training inputs.  We quantify this
relationship in Fig.~\ref{fig:howitworks}.  From the machine learning
perspective, overfitting is harmful because it produces models that lack
predictive power.  In this paper, we show another harm of overfitting:
the leakage of sensitive information about the training data.

As we explained in Section~\ref{sec:evaluation}, overfitting is not the
only reason why our inference attacks work.  Different machine learning
models, due to their different structures, ``remember'' different amounts
of information about their training datasets.  This leads to different
amounts of information leakage even if the models are overfitted to the
same degree (see Table~\ref{table:target-accuracy}).

\section{Mitigation}
\label{sec:mitigation}

As explained in Section~\ref{sec:whyattackworks}, overfitting is an
important (but not the only) reason why machine learning models leak
information about their training datasets.  Of course, overfitting is a
canonical problem in machine learning because it limits the predictive
power and generalizability of models.  This means that instead of
the usual tradeoff between utility and privacy, machine learning
research and privacy research have similar objectives in this case.
Regularization techniques such as dropout~\cite{srivastava2014dropout}
can help defeat overfitting and also strengthen privacy guarantees
in neural networks~\cite{jain2015drop}.  Regularization is also
used for objective perturbation in differentially private machine
learning~\cite{chaudhuri2011differentially}.

(Ideal) well-regularized models should not leak much information about
their training data, and our attack can serve as a metric to quantify
this.  Also, models with a trivial structure (e.g., XOR of some input
features) generalize to the entire universe and do not leak information.

If the training process is differentially
private~\cite{dwork2011differential}, the probability of producing a
given model from a training dataset that includes a particular record is
close to the probability of producing the same model when this record
is not included.  Differentially private models are, by construction,
secure against membership inference attacks of the kind developed in this
paper because our attacks operate solely on the outputs of the model,
without any auxiliary information.  One obstacle is that differentially
private models may significantly reduce the model's prediction accuracy
for small $\epsilon$ values.  In Section~\ref{sec:relatedwork}, we survey
some of the related work in this area.

In the case of machine learning as a service, platform operators such as
Google and Amazon have significant responsibility to the users of their
services.  In their current form, these services simply accept the data,
produce a model of unknown type and structure, and return an opaque API to
this model that data owners use as they see fit, without any understanding
that by doing so, they may be leaking out their data.  Machine learning
services do not inform their customers about the risks of overfitting
or the harm that may result from models trained on inadequate datasets
(for example, with unrepresentative records or too few representatives
for certain classes).

Instead, when adaptively choosing a model for a customer-supplied dataset,
services such as Google Prediction API and Amazon ML should take into
account not only the accuracy of the model but also the risk that it
will leak information about its training data.  Furthermore, they need
to explicitly warn customers about this risk and provide more visibility
into the model and the methods that can be used to reduce this leakage.
Our inference attacks can be used as metrics to quantify leakage from a
specific model, and also to measure the effectiveness of future privacy
protection techniques deployed by machine-learning services.

\subsection{Mitigation strategies}

We quantitatively evaluate several defenses against membership inference.

\paragraphbe{Restrict the prediction vector to top $k$ classes.}
When the number of classes is large, many classes may have very small
probabilities in the model's prediction vector.  The model will still
be useful if it only outputs the probabilities of the most likely
$k$ classes.  To implement this, we add a filter to the last layer of
the model.  The smaller $k$ is, the less information the model leaks.
In the extreme case, the model returns only the label of the most likely
class without reporting its probability.

\paragraphbe{Coarsen precision of the prediction vector.}
To implement this, we round the classification probabilities in the
prediction vector down to $d$ floating point digits.  The smaller $d$
is, the less information the model leaks.

\paragraphbe{Increase entropy of the prediction vector.}
One of the signals that membership inference exploits is the difference
between the prediction entropy of the target model on its training inputs
versus other inputs.  As a mitigation technique for neural-network
models, we can modify (or add) the softmax layer and increase its
normalizing temperature $t > 0$.  The softmax layer converts the logits
computed for each class into probabilities.  For the logits vector
$\mathbf{z}$, the $i^{\mathrm{th}}$ output of the softmax function with
temperature $t$ is $\frac{e^{z_i/t}}{\sum_j e^{z_j/t}}.$ This technique,
also used in knowledge distillation and information transfer between
models~\cite{hinton2015distilling}, would increase the entropy of the
prediction vector.  Note that for a very large temperature, the output
becomes almost uniform and independent of its input, thus leaking no
information.

\paragraphbe{Use regularization.}
Regularization techniques are used to overcome overfitting in machine
learning.  We use $L_2$-norm standard regularization that penalizes large
parameters by adding $\lambda \sum_i \theta_i^2$ to the model's loss
function, where $\theta_i$s are model's parameters.  We implement this
technique with various values for the regularization factor $\lambda$.
The larger $\lambda$ is, the stronger the effect of regularization during
the training.

\begin{table}
\begin{center}
\begin{tabular}{l | c c c c}
{\bf Purchase dataset} & {\em Testing} & {\em Attack} & {\em Attack} & {\em Attack} \\
& {\em Accuracy} & {\em Total Accuracy} & {\em Precision} & {\em Recall} \\\hline
  No Mitigation					& 0.66 & 0.92 & 0.87 & 1.00 \\\hline
  Top $k=3$						& 0.66 & 0.92 & 0.87 & 0.99 \\
  Top $k=1$						& 0.66 & 0.89 & 0.83 & 1.00 \\
  Top $k=1$ label				& 0.66 & 0.66 & 0.60 & 0.99 \\\hline
  Rounding $d=3$				& 0.66 & 0.92 & 0.87 & 0.99 \\
  Rounding $d=1$				& 0.66 & 0.89 & 0.83 & 1.00 \\\hline
  Temperature $t=5$				& 0.66 & 0.88 & 0.86 & 0.93 \\
  Temperature $t=20$			& 0.66 & 0.84 & 0.83 & 0.86 \\\hline
  L2 $\lambda=1e-4$				& 0.68 & 0.87 & 0.81 & 0.96 \\
  L2 $\lambda=1e-3$				& 0.72 & 0.77 & 0.73 & 0.86 \\  
  L2 $\lambda=1e-2$				& 0.63 & 0.53 & 0.54 & 0.52 \\\hline  
\end{tabular}
\\[10pt]
\begin{tabular}{l | c c c c}
{\bf Hospital dataset} & {\em Testing} & {\em Attack} & {\em Attack} & {\em Attack} \\
& {\em Accuracy} & {\em Total Accuracy} & {\em Precision} & {\em Recall} \\\hline
  No Mitigation					& 0.55 & 0.83 & 0.77 & 0.95 \\\hline
  Top $k=3$						& 0.55 & 0.83 & 0.77 & 0.95 \\
  Top $k=1$						& 0.55 & 0.82 & 0.76 & 0.95 \\
  Top $k=1$ label				& 0.55 & 0.73 & 0.67 & 0.93 \\\hline
  Rounding $d=3$				& 0.55 & 0.83 & 0.77 & 0.95 \\
  Rounding $d=1$				& 0.55 & 0.81 & 0.75 & 0.96 \\\hline
  Temperature $t=5$				& 0.55 & 0.79 & 0.77 & 0.83 \\
  Temperature $t=20$			& 0.55 & 0.76 & 0.76 & 0.76 \\\hline
  L2 $\lambda=1e-4$				& 0.56 & 0.80 & 0.74 & 0.92 \\
  L2 $\lambda=5e-4$				& 0.57 & 0.73 & 0.69 & 0.86 \\  
  L2 $\lambda=1e-3$				& 0.56 & 0.66 & 0.64 & 0.73 \\
  L2 $\lambda=5e-3$				& 0.35 & 0.52 & 0.52 & 0.53 \\\hline  
\end{tabular}
\end{center}
\caption{The accuracy of the target models with different mitigation
techniques on the purchase and Texas hospital-stay datasets (both with
100 classes), as well as total accuracy, precision, and recall of the
membership inference attack.  The relative reduction in the metrics for
the attack shows the effectiveness of the mitigation strategy.}
\label{table:mitigation} \end{table}

\subsection{Evaluation of mitigation strategies}

To evaluate the effectiveness of different mitigation strategies,
we implemented all of them in locally trained models over which we
have full control.  The inference attack, however, still assumes only
black-box access to the resulting models.  The baseline model for these
experiments is a neural network with one hidden layer with 256 units
(for the purchase dataset) and 1,000 units (for the Texas hospital-stay
dataset).  We use $\mathtt{Tanh}$ as the activation function.

Table~\ref{table:mitigation} shows the results of our evaluation.
It compares different mitigation strategies based on how they degrade the
accuracy of our attack relative to the attack on a model that does not
use any mitigation.  The mitigation strategies that we implemented did
not impose any cost on the target model's prediction accuracy, and in the
case of regularization, the target model's prediction accuracy increased
as expected.  Note that more regularization (by increasing $\lambda$
even further) would potentially result in a significant reduction of
the target model's test accuracy, even if it foils membership inference.
This is shown in the table for $\lambda=1e-2$ on the purchase dataset,
and for $\lambda=5e-3$ on the Texas hospital stay dataset.

Overall, our attack is robust against these mitigation strategies.
Filtering out low-probability classes from the prediction vector and
limiting the vector to the top 1 or 3 most likely classes does not
foil the attack.  Even \textbf{restricting the prediction vector to a
single label (most likely class), which is the absolute minimum a model
must output to remain useful, is not enough to fully prevent membership
inference}.  Our attack can still exploit the \emph{mislabeling behavior}
of the target model because members and non-members of the training
dataset are mislabeled differently (assigned to different wrong classes).
If the prediction vector contains probabilities in addition to the
labels, the model leaks even more information that can be used for
membership inference.

Some of the mitigation methods are not suitable for
machine-learning-as-service APIs used by general applications and
services.  Regularization, however, appears to be necessary and useful.
As mentioned above, it (1) generalizes the model and improves its
predictive power and (2) decreases the model's information leakage about
its training dataset.  However, regularization needs to be deployed
carefully to avoid damaging the model's performance on the test datasets.

\section{Related Work}
\label{sec:relatedwork}

\paragraphbe{Attacks on statistical and machine learning models.}
In \cite{ateniese2015hacking}, knowledge of the parameters of SVM and
HMM models is used to infer general statistical information about the
training dataset, for example, whether records of a particular race
were used during training.  By contrast, our inference attacks work in a
black-box setting, without any knowledge of the model's parameters, and
infer information about \emph{specific records} in the training dataset,
as opposed to general statistics.

Homer et al.~\cite{homer2008resolving} developed a technique, which
was further studied in~\cite{dwork2015robust, backes2016membership},
for inferring the presence of a particular genome in a dataset, based
on comparing the published statistics about this dataset (in particular,
minor allele frequencies) to the distribution of these statistics in the
general population.  By contrast, our inference attacks target trained
machine learning models, not explicit statistics.

Other attacks on machine learning include~\cite{calandrino}, where the
adversary exploits \emph{changes} in the outputs of a collaborative
recommender system to infer inputs that caused these changes.  These
attacks exploit temporal behavior specific to the recommender systems
based on collaborative filtering.

\begin{figure}[t]
\centering
\includegraphics[width=0.19\columnwidth]{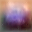}
\includegraphics[width=0.19\columnwidth]{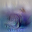}
\includegraphics[width=0.19\columnwidth]{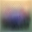}
\includegraphics[width=0.19\columnwidth]{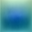}
\includegraphics[width=0.19\columnwidth]{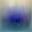}\\[3pt]
\includegraphics[width=0.19\columnwidth]{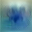}
\includegraphics[width=0.19\columnwidth]{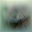}
\includegraphics[width=0.19\columnwidth]{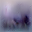}
\includegraphics[width=0.19\columnwidth]{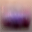}
\includegraphics[width=0.19\columnwidth]{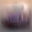}
\caption{Images produced by model inversion on a trained CIFAR-10 model.
Top: airplane, automobile, bird, cat, deer.  Bottom: dog, frog, horse,
ship, truck.  The images do not correspond to any specific image from
the training dataset, are not human-recognizable, and at best (e.g.,
the truck class image) are vaguely similar to the average image of all
objects in a given class.}
\label{fig:modelinversion}
\vspace{-2ex}
\end{figure}

\paragraphbe{Model inversion.}
Model inversion~\cite{fredrikson2014privacy, fredrikson2015model} uses
the output of a model applied to a hidden input to infer certain features
of this input.  See~\cite{frankblog} for a detailed analysis of this
attack and an explanation of why it does not necessarily entail a privacy
breach.  For example, in the specific case of pharmacogenetics analyzed
in~\cite{fredrikson2014privacy}, the model captures the correlation
between the patient's genotype and the dosage of a certain medicine.
This correlation is a valid scientific fact that holds for all patients,
regardless of whether they were included in the model's training dataset
or not.  It is not possible to prevent disclosure due to population
statistics~\cite{dworknaor}.

In general, model inversion cannot tell whether a particular record
was used as part of the model's training dataset.  Given a record and
a model, model inversion works exactly the same way when the record
was used to train the model and when it was not used.  In the case of
pharmacogenetics~\cite{fredrikson2014privacy}, model inversion produces
almost identical results for members and non-members.  Due to the
overfitting of the model, the results are a little (4\%) more accurate
for the members, but this accuracy can only be measured in retrospect,
if the adversary already knows the ground truth (i.e., which records
are indeed members of the model's training dataset).  By contrast, our
goal is to construct a decision procedure that distinguishes members
from non-members.

Model inversion has also been applied to face recognition
models~\cite{fredrikson2015model}.  In this scenario, the model's output
is set to $1$ for class $i$ and $0$ for the rest, and model inversion
is used to construct an input that produces these outputs.  This input
is not an actual member of the training dataset but simply an average
of the features that ``characterize'' the class.

In the face recognition scenario\textemdash and \emph{only} in this
specific scenario\textemdash each output class of the model is associated
with a single person.  All training images for this class are different
photos of that person, thus model inversion constructs an artificial
image that is an average of these photos.  Because they all depict the
same person, this average is recognizable (by a human) as that person.
Critically, model inversion does not produce any \emph{specific image}
from the training dataset, which is the definition of membership
inference.

If the images in a class are diverse (e.g., if the class contains multiple
individuals or many different objects), the results of model inversion
as used in~\cite{fredrikson2015model} are semantically meaningless
and not recognizable as any specific image from the training dataset.
To illustrate this, we ran model inversion against a convolutional neural
network\footnote{\url{https://github.com/Lasagne/Recipes/blob/master/modelzoo/cifar10_nin.py}}
trained on the CIFAR-10 dataset, which is a standard benchmark for object
recognition models.  Each class includes different images of a single
type of object (e.g., an airplane).  Figure~\ref{fig:modelinversion}
shows the images ``reconstructed'' by model inversion.  As expected,
they do not depict any recognizable object, let alone an image from the
training dataset.  We expect similar results for other models, too.
For the pharmacogenetics model mentioned above, this form of model
inversion produces an average of different patients' genomes.  For the
model that classifies location traces into geosocial profiles (see
Section~\ref{locations}), it produces an average of the location traces
of different people.  In both cases, the results of model inversion are
not associated with any specific individual or specific training input.

In summary, model inversion produces the average of the features that at
best can characterize an entire output class.  It does not (1) construct
a specific member of the training dataset, nor (2) given an input and
a model, determines if this specific input was used to train the model.

\paragraphbe{Model extraction.}
Model extraction attacks~\cite{tramer2016stealing} aim to extract the
parameters of a model trained on private data.  The attacker's goal is
to construct a model whose predictive performance on validation data is
similar to the target model.

Model extraction can be a stepping stone for inferring information
about the model's training dataset.  In~\cite{tramer2016stealing}, this
is illustrated for a specific type of models called kernel logistic
regression (KLR)~\cite{zhu2001kernel}.  In KLR models, the kernel
function includes a tiny fraction of the training data (so called
``import points'') directly into the model.  Since import points are
parameters of the model, extracting them results in the leakage of that
particular part of the data.  This result is very specific to KLR and
does not extend to other types of models since they do not explicitly
store training data in their parameters.

Even for KLR models, leakage is not quantified other than via visual
similarity of a few chosen import points and ``the closest (in L1 norm)
extracted representers'' on the MNIST dataset of handwritten digits.
In MNIST, all members of a class are very similar (e.g., all members
of the first class are different ways of writing digit ``1'').  Thus,
any extracted digit must be similar to all images in its class, whether
this digit was in the training set or not.

\paragraphbe{Privacy-preserving machine learning.}
Existing literature on privacy protection in machine learning
focuses mostly on how to learn without direct access to the
training data.  Secure multiparty computation (SMC) has been
used for learning decision trees~\cite{lindell2000privacy},
linear regression functions~\cite{du2004privacy},
Naive Bayes classifiers~\cite{vaidya2008privacy}, and k-means
clustering~\cite{jagannathan2005privacy}.  The goal is to limit
information leakage during training.  The training algorithm is the same
as in the non-privacy-preserving case, thus the resulting models are as
vulnerable to inference attacks as any conventionally trained model.
This also holds for the models trained by computing on encrypted
data~\cite{bos2014private,barni2011privacy,xie2014crypto}.

Differential privacy~\cite{dwork2011differential}
has been applied to linear and logistic
regression~\cite{zhang2012functional, chaudhuri2009privacy},
support vector machines~\cite{rubinstein2012learning},
risk minimization~\cite{chaudhuri2011differentially,
wainwright2012privacy, bassily2014private}, deep
learning~\cite{shokri2015privacy, abadi2016deep}, learning an
unknown probability distribution over a discrete population from
random samples~\cite{diakonikolas2015differentially}, and releasing
hyper-parameters and classifier accuracy~\cite{kusner2015differentially}.
By definition, differentially private models limit the success probability
of membership inference attacks based solely on the model, which includes
the attacks described in this paper.

\section{Conclusions}

We have designed, implemented, and evaluated the first membership
inference attack against machine learning models, notably black-box
models trained in the cloud using Google Prediction API and Amazon ML.
Our attack is a general, quantitative approach to understanding how
machine learning models leak information about their training datasets.
When choosing the type of the model to train or a machine learning
service to use, our attack can be used as one of the selection metrics.

Our key technical innovation is the shadow training technique that trains
an attack model to distinguish the target model's outputs on members
versus non-members of its training dataset.  We demonstrate that shadow
models used in this attack can be effectively created using synthetic
or noisy data.  In the case of synthetic data generated from the target
model itself, the attack does not require any prior knowledge about the
distribution of the target model's training data.

Membership in hospital-stay and other health-care datasets is sensitive
from the privacy perspective.  Therefore, our results have substantial
practical privacy implications.

\paragraphbe{Acknowledgments.}
Thanks to Adam Smith for explaining differential privacy and the state
of the art in membership inference attacks based on explicit statistics.

This work was supported by the NSF grant 1409442 and a Google Research
Award.

\bibliographystyle{IEEEtranS}
\balance

\end{document}